\documentclass{article}
\vspace{.5cm}

\usepackage{amssymb,latexsym}

\vspace{.5cm}
\vspace{.5cm}
\usepackage[pdftex]{graphicx}
\vspace{.5cm}
\usepackage{hyperref}
\usepackage{xcolor}
\hypersetup{
    colorlinks,
    linkcolor={red!50!black},
    citecolor={blue!50!black},
    urlcolor={blue!80!black}
}
\vspace{.5cm}

\vspace{.5cm}

\begin{document}
\vspace{.5cm}

\pagestyle{plain}
\vspace{.5cm}

\newtheorem{theorem}{Theorem}[section]

\newtheorem{proposition}[theorem]{Proposition}
\vspace{.5cm}
\newtheorem{lemma}[theorem]{Lemma}
\vspace{.5cm}
\newtheorem{corollary}[theorem]{Corollary}
\vspace{.5cm}
\newtheorem{definition}[theorem]{Definition}
\vspace{.5cm}
\newtheorem{remark}[theorem]{Remark}
\vspace{.5cm}
\newtheorem{exempl}{Example}[section]

\newenvironment{example}{\begin{exempl}  \em}{\hfill $\square$

\end{exempl}}  \vspace{.5cm}
\vspace{.5cm}

\renewcommand{\contentsname}{ }
\vspace{.5cm}

\title{Graph rewrites, from graphic lambda calculus, to chemlambda, to directed interaction combinators}
\vspace{.5cm}
\author{Marius Buliga \\ 
\\
Institute of Mathematics, Romanian Academy \\
P.O. BOX 1-764, RO 014700\\
Bucure\c sti, Romania\\ 
{\footnotesize Marius.Buliga@imar.ro , mbuliga@protonmail.ch}}  \vspace{.5cm}
\vspace{.5cm}

\date{07.07.2020}
\vspace{.5cm}

\maketitle

\begin{abstract}
Here I report about the modifications of and relations between graphic lambda calculus,  various formalisms which appeared under the name chemlambda and a version of directed interaction combinators.

This is part of the study and experiments with the artificial chemistry chemlambda and the relations with lambda calculus or interaction combinators, as described in \href{https://arxiv.org/abs/2003.14332}{arXiv:2003.14332} and available from the entry page \href{https://chemlambda.github.io/index.html}{chemlambda.github.io} \cite{entry}.
\end{abstract}
\vspace{.5cm}

\section*{Context}
The chemlambda project context and relevant previous work are explained in Section 4 (About this project) of \cite{buligachemlambda} \href{https://arxiv.org/abs/2003.14332}{arXiv:2003.14332}  See also, for more mathematical background, the presentation \cite{buliganovo} \href{https://mbuliga.github.io/novo/presentation.html}{Emergent rewrites in knot theory and logic}.

 Here I report about the modifications of the various formalisms which are related to chemlambda.   The online varsion of this article, which will be kept up to date, is \href{https://mbuliga.github.io/quinegraphs/history-of-chemlambda.html}{Graph rewrites, from emergent algebras to chemlambda}. See also the site of \href{https://chemlambda.github.io/index.html}{all chemlambda projects}.

\tableofcontents

\section{Graphic lambda calculus}
\label{GraphicLambdaCalculus}
Introduced in \cite{buligaglc}  \href{https://arxiv.org/abs/1305.5786}{arXiv:1305.5786}, graphic lambda calculus  has the following doupbel push-out (DPO \cite{dpo}) graph rewrites:
\vspace{.5cm}
 
\centerline{\includegraphics[width=0.75\textwidth]{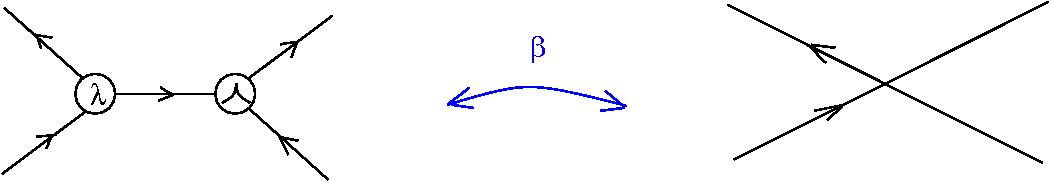}
}
\vspace{.5cm}

\vspace{.5cm}
\centerline{\includegraphics[width=0.5\textwidth]{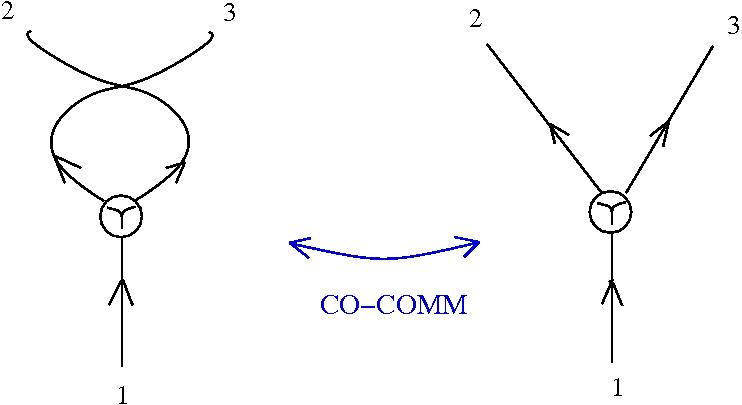}
}
\vspace{.5cm}

\vspace{.5cm}
 
\centerline{\includegraphics[width=0.6\textwidth]{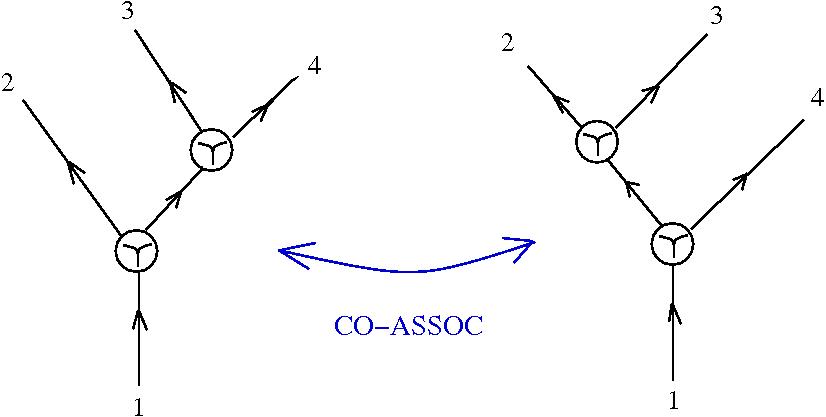}
}
\vspace{.5cm}

\centerline{\includegraphics[width=0.6\textwidth]{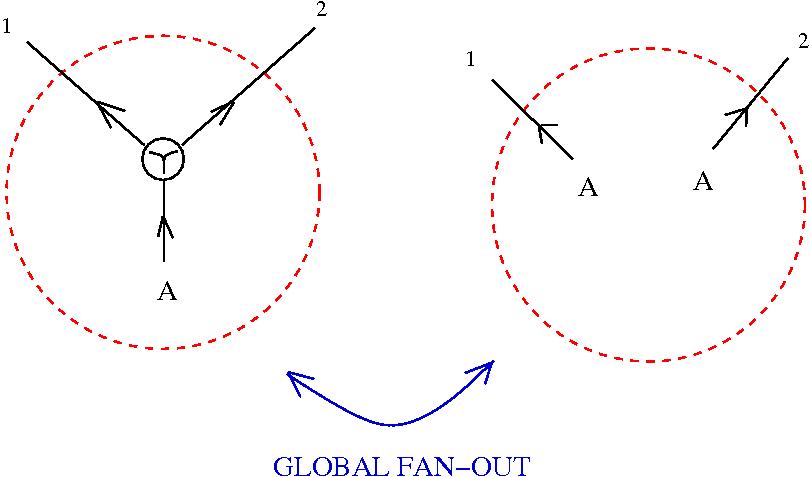}
}
\vspace{.5cm}

\vspace{.5cm}
 
\centerline{\includegraphics[width=0.5\textwidth]{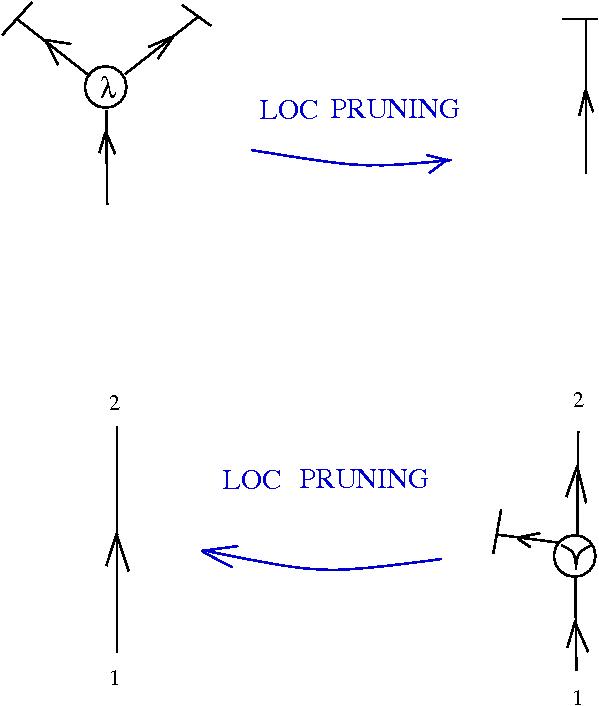}
}
\vspace{.5cm}

\centerline{\includegraphics[width=0.6\textwidth]{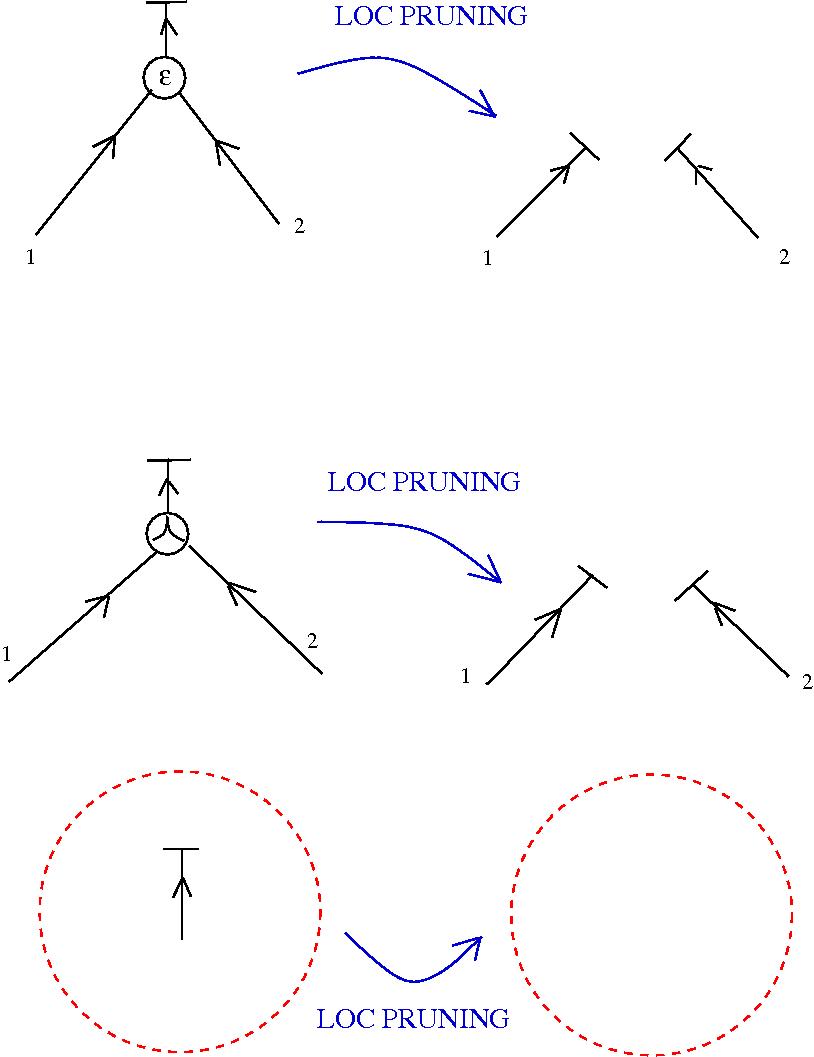}
}
\vspace{.5cm}

\vspace{.5cm}
 
\centerline{\includegraphics[width=0.6\textwidth]{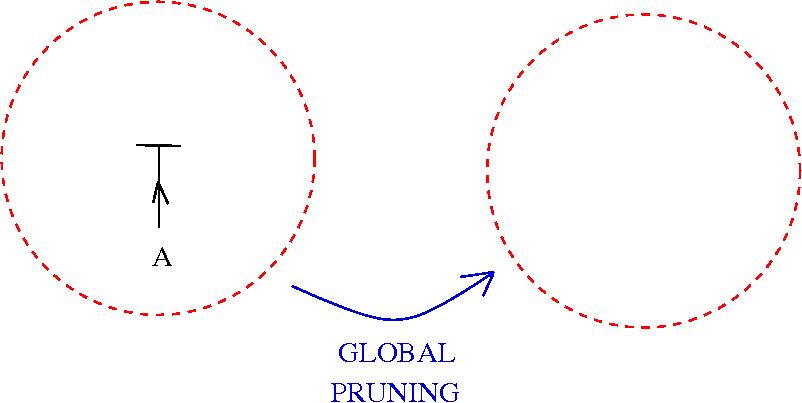}
}
\vspace{.5cm}

\vspace{.5cm}

\centerline{\includegraphics[width=0.6\textwidth]{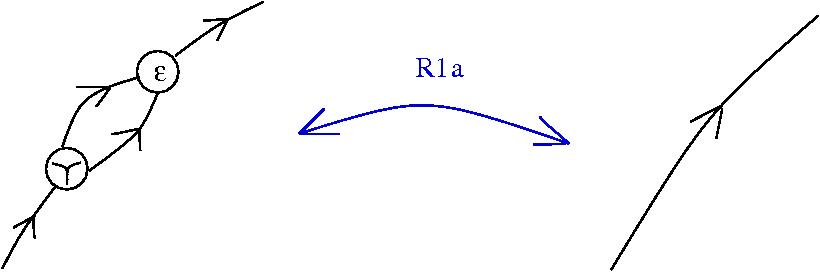}
}
\vspace{.5cm}

\vspace{.5cm}
 
\centerline{\includegraphics[width=0.6\textwidth]{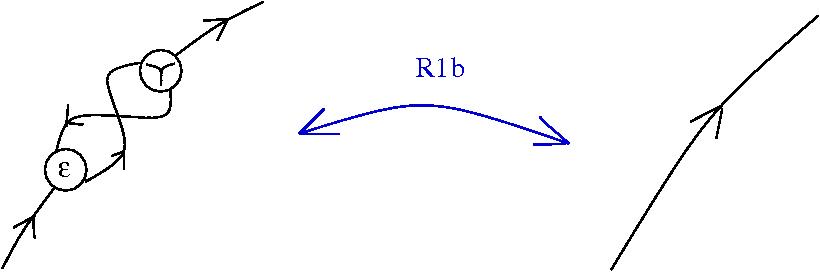}
}
\vspace{.5cm}

\centerline{\includegraphics[width=0.6\textwidth]{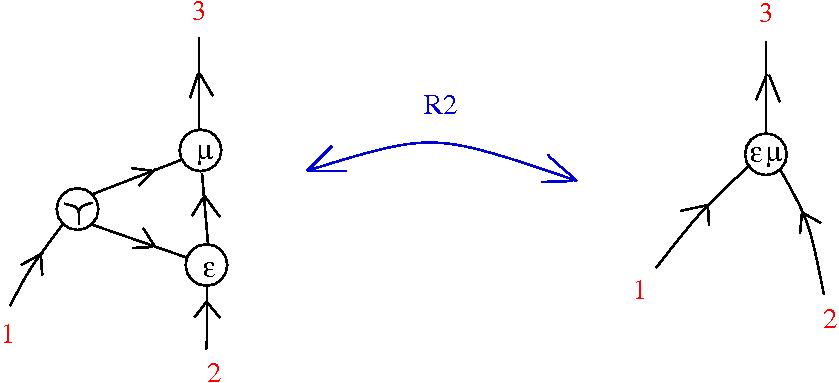}
}
\vspace{.5cm}

\vspace{.5cm}
 
\centerline{\includegraphics[width=0.6\textwidth]{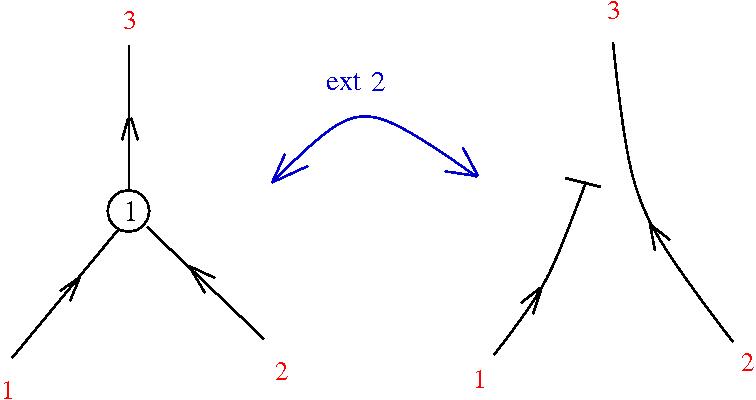}
}
\vspace{.5cm}
Graphic lambda calculus does not have an algorithm of application of rewrites. 

There is an algorithm for conversion of untyped lambda terms into glc graphs. Here are two examples of reductions:
\begin{enumerate}
\item[-] reduction of the Omega combinator

\vspace{.5cm}
 
\centerline{\includegraphics[width=0.5\textwidth]{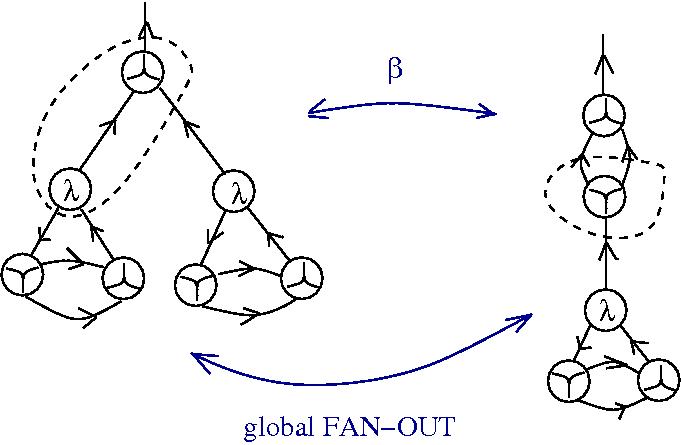}
}
\vspace{.5cm}
\item[-] reduction of the term SKK

\vspace{.5cm}
 
\centerline{\includegraphics[width=0.9\textwidth]{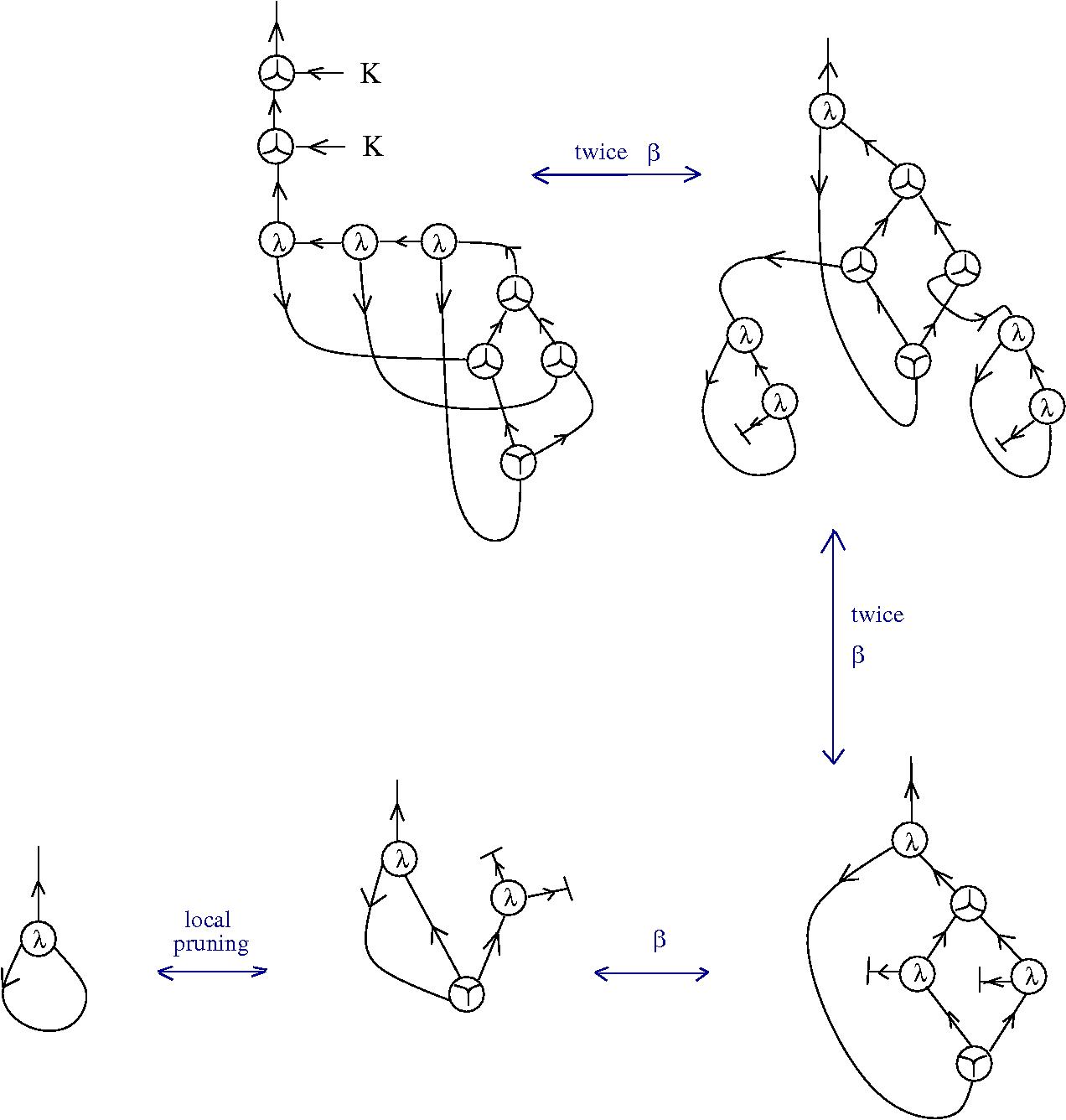}
}
\end{enumerate}
\vspace{.5cm}
Graphic lambda calculus is interesting because it can also represent the graphical version of emergent algebras via decorated tangle diagrams explained in \href{https://arxiv.org/abs/1103.6007}{arXiv:1103.6007}. It can therefore be used to represent and compute (reduce) differential calculus in metric spaces endowed with dilation structures \href{https://arxiv.org/abs/0810.5042}{arXiv:0810.5042}.

 This is done via the identification of decorated crossings as
\vspace{.5cm}
 
\centerline{\includegraphics[width=0.6\textwidth]{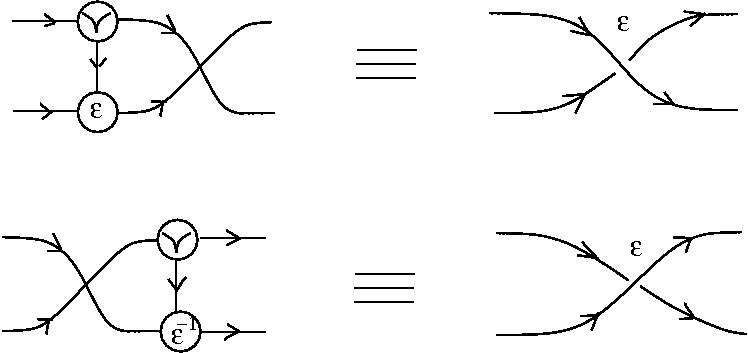}
}
\vspace{.5cm}

Another, different identification of undecorated crossings, which can be done with GLC, uses the application and abstraction nodes

\vspace{.5cm}
 
\centerline{\includegraphics[width=0.6\textwidth]{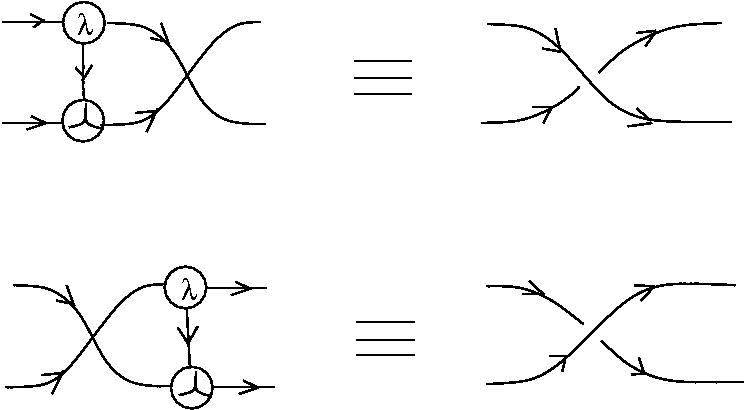}
}
\vspace{.5cm}

 Under this identifications the equivalent of the beta rewrite in terms of tangle diagrams becomes equivalent with the "splice" and "loop" graph rewrites.

\vspace{.5cm}
 
\centerline{\includegraphics[width=0.7\textwidth]{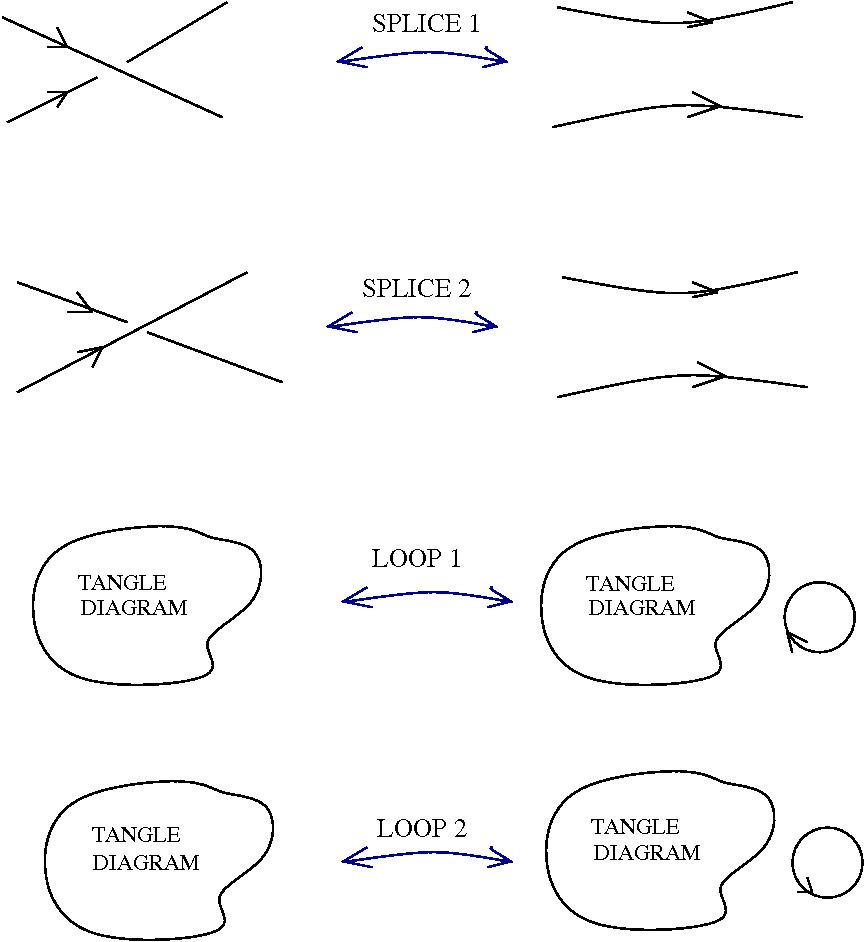}
}

\vspace{.5cm}

Both these identifications apply for a notion of tangle diagrams which differs from the usual one. Namely we use tangle diagrams which are not constrained to be planar graphs. 
More precisely, here a tangle diagram is an oriented \href{http://citeseerx.ist.psu.edu/viewdoc/download?doi=10.1.1.18.5446&rep=rep1&type=pdf}{port graph in the sense of Bawden} \cite{bawden1} \cite{bawden2}, with nodes which are 4-valent and oriented links. The nodes can be either undecorated, i.e. of the two kinds of oriented crossings, or decorated, i.e. the two kinds of decorated crossings are supplementary decorated with an element of a commutative group.

The difference from usual (oriented and with decorated crossings) tangle diagrams is that as graphs these diagrams are not supposed to be planar. By Kuratowski theorem, a graph is planar if it does not contain certain local patterns. In the terms of local machines explained in \cite{buligaalife} 
\href{https://mbuliga.github.io/quinegraphs/ic-vs-chem.html#icvschem}{Alife properties of directed interaction combinators vs. chemlambda} 
we can't detect with a local machine the absence of a local pattern in a graph, only the presence of it.

 In joint work with L.H. Kauffman  \cite{buligakauffmanglc} \href{https://arxiv.org/abs/1312.4333}{arXiv:1312.4333}, in section 5 Kauffman calls for a topological version of GLC, based exclusively on manipulations of tangle diagrams.

\section{Chemlambda v1 (chemical concrete machine)}
\label{ChemlambdaV1}
Introduced in \cite{buligachem} \href{https://arxiv.org/abs/1309.6914}{arXiv:1309.6914}, chemlambda v1 or the chemical concrete machine is a modification of GLC which contains only local rewrites. It also proposes for the first time to see the graphs as molecules and the rewrites as chemical reactions mediated by rewriting enzymes. Beyond that, chemlambda v1 still lacks a clearly defined algorithm of rewriting.

The elements of chemlambda v1 are

\vspace{.5cm}
 
\centerline{\includegraphics[width=0.6\textwidth]{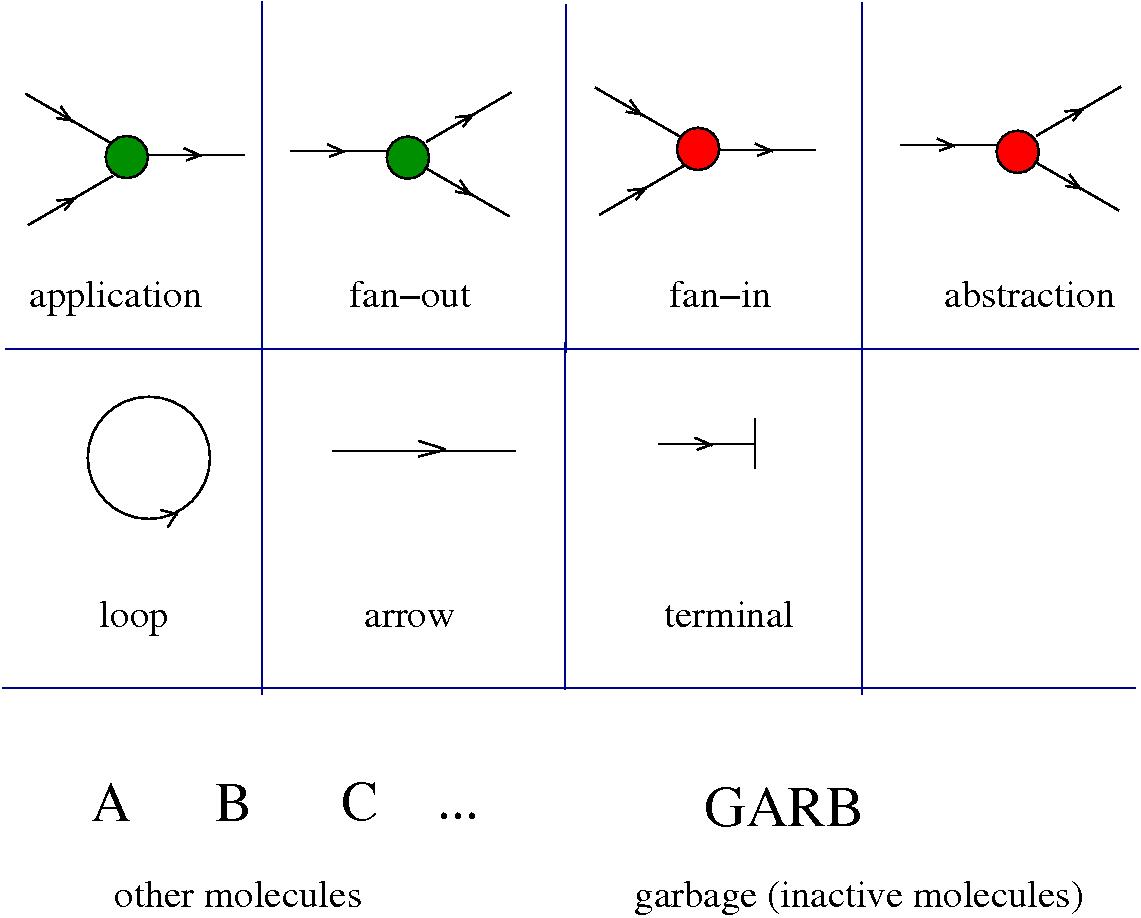}
}
\centerline{\includegraphics[width=0.5\textwidth]{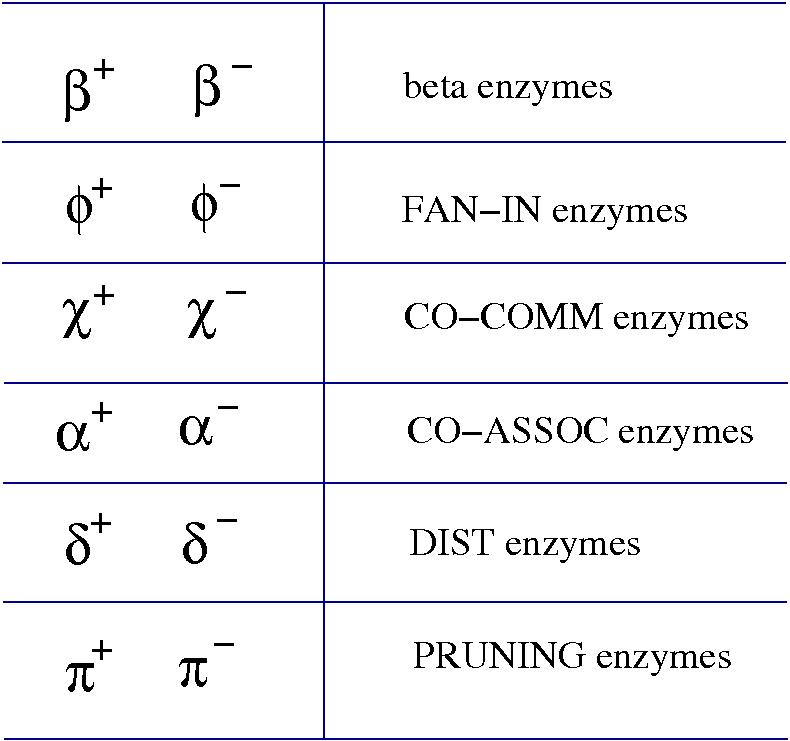}
}
\vspace{.5cm}
The graphs are called molecules and the rewrites are mediated by enzymes.

\vspace{.5cm}
 
\centerline{\includegraphics[width=0.9\textwidth]{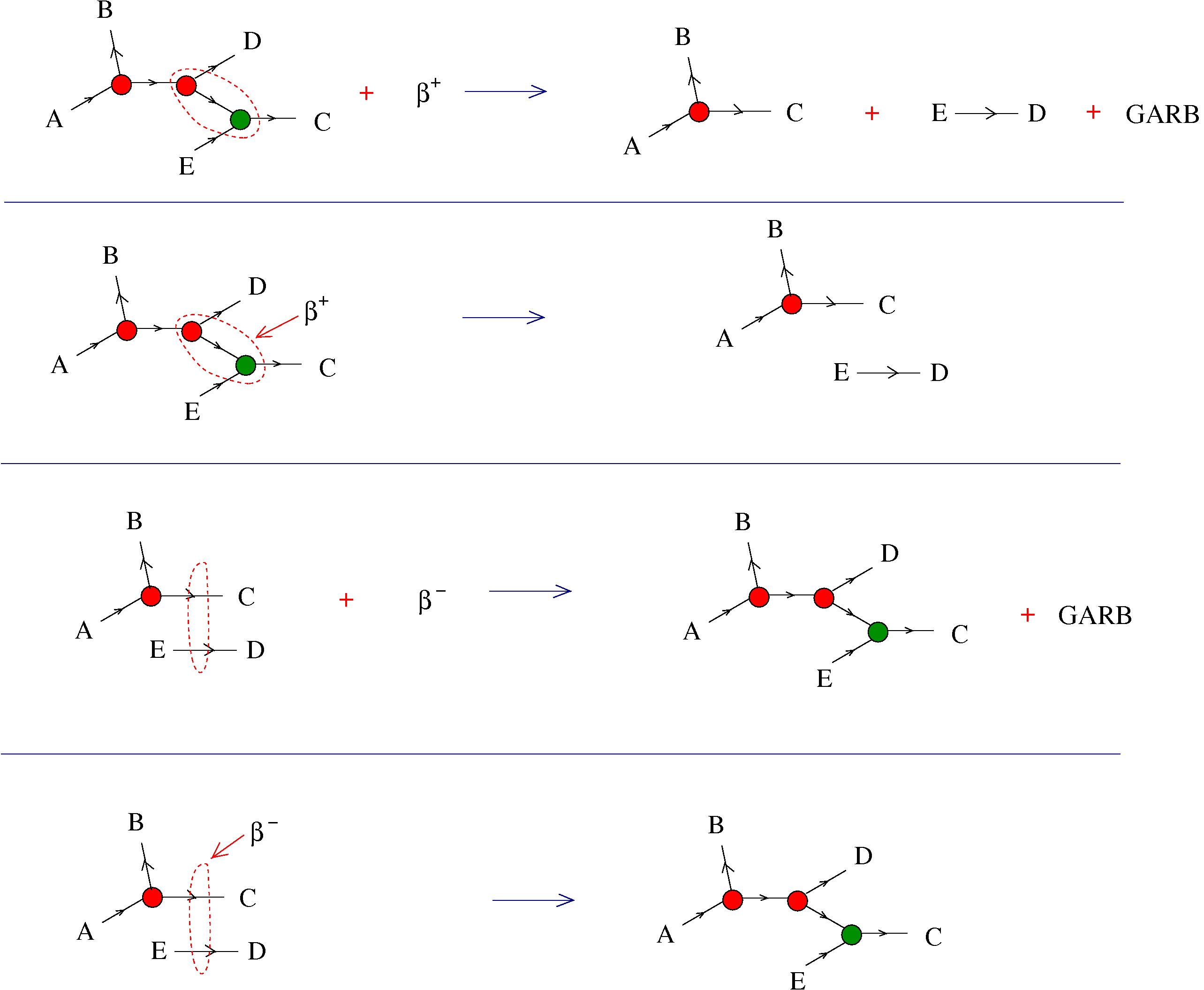}
}
\vspace{.5cm}

The elements of GLC as translated into chemlambda are:

\vspace{.5cm}
 
\centerline{\includegraphics[width=0.7\textwidth]{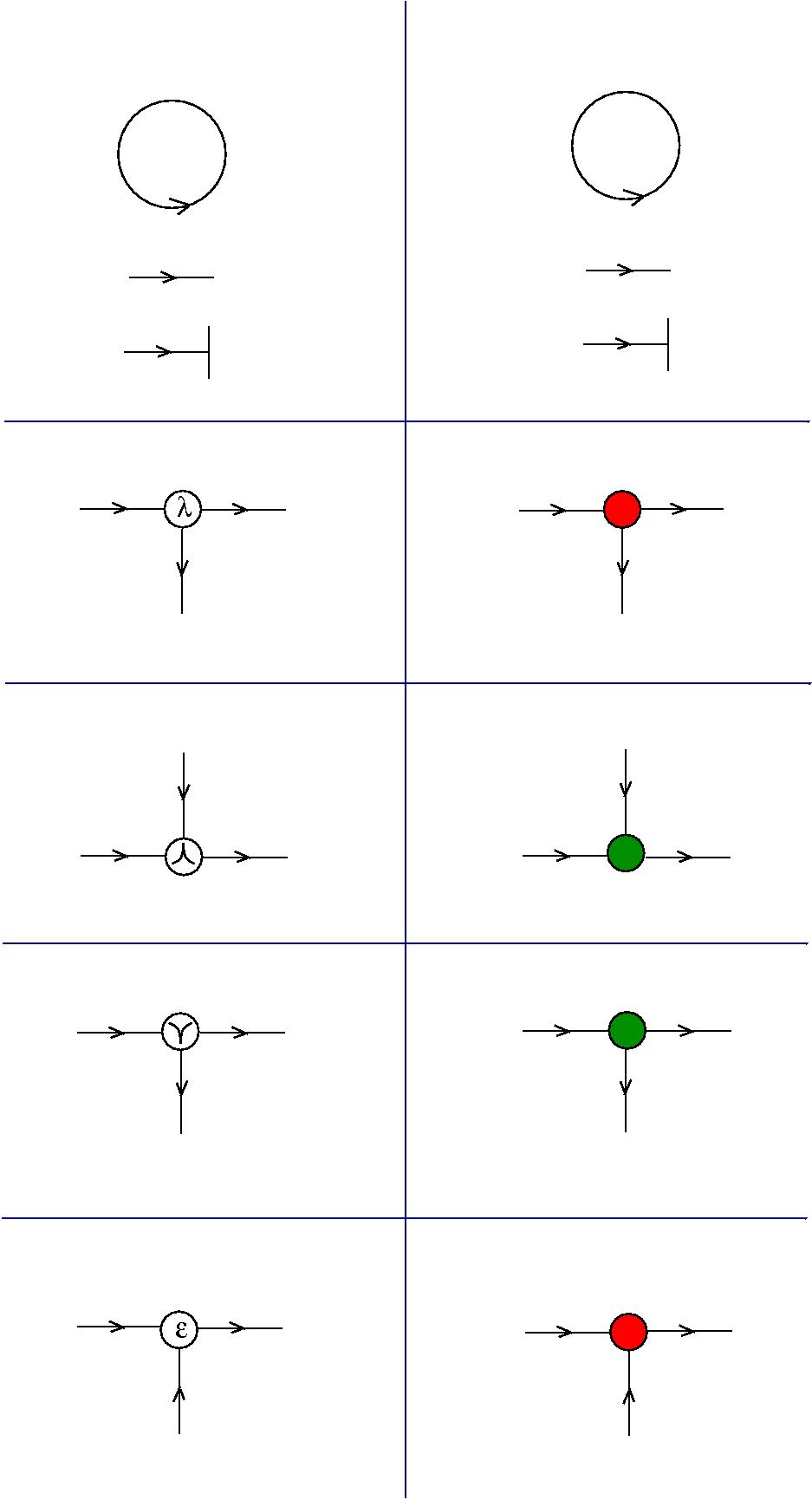}
}
\vspace{.5cm}

The rewrites are:

\vspace{.5cm}
 
\centerline{\includegraphics[width=0.9\textwidth]{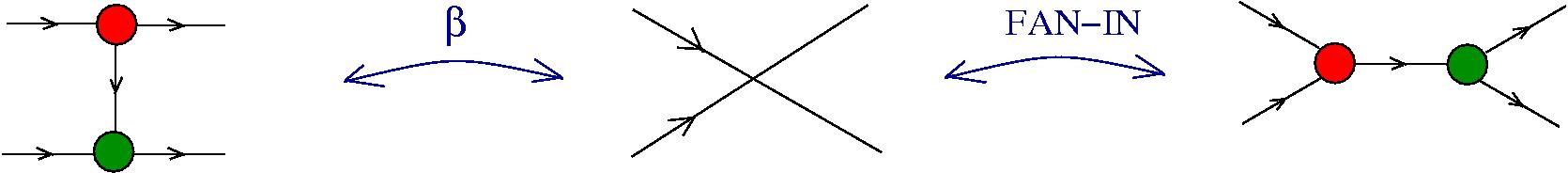}
}
\vspace{.5cm}

\centerline{\includegraphics[width=0.7\textwidth]{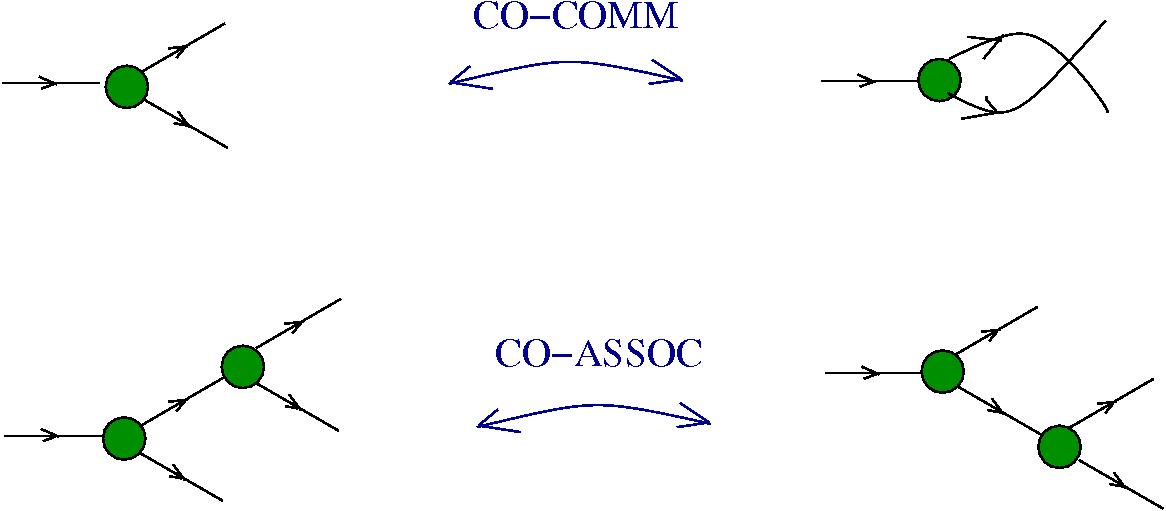}
}
\vspace{.5cm}

\vspace{.5cm}
 
\centerline{\includegraphics[width=0.7\textwidth]{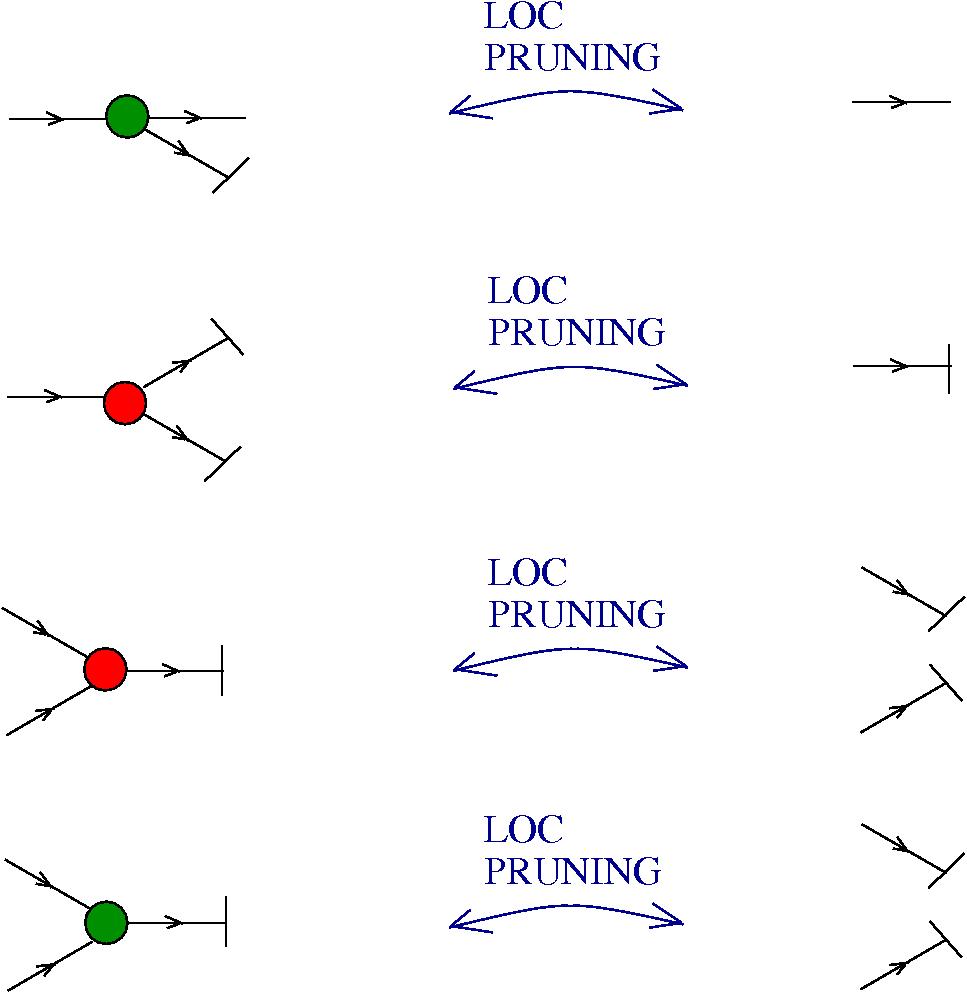}
}
\vspace{.5cm}

\vspace{.5cm}

\centerline{\includegraphics[width=0.4\textwidth]{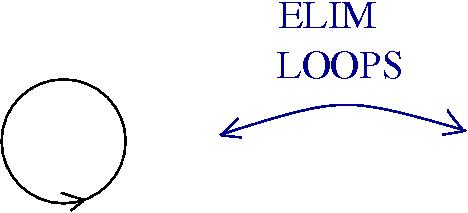}
}
\vspace{.5cm}

\vspace{.5cm}
 
\centerline{\includegraphics[width=0.7\textwidth]{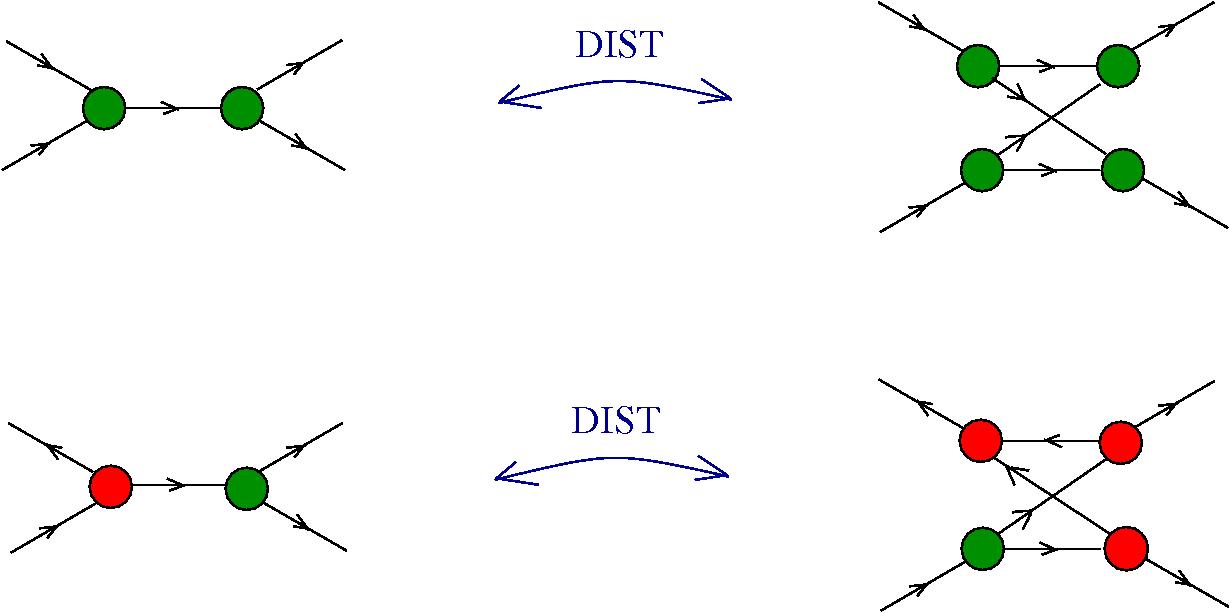}
}
\vspace{.5cm}

With these rewrites we can do some lambda calculus reductions, by using the same parsing of lambda calculus terms to (chemlambda) graphs as in GLC. For example the ENELSE lambda term.

\vspace{.5cm}
 
\centerline{\includegraphics[width=0.5\textwidth]{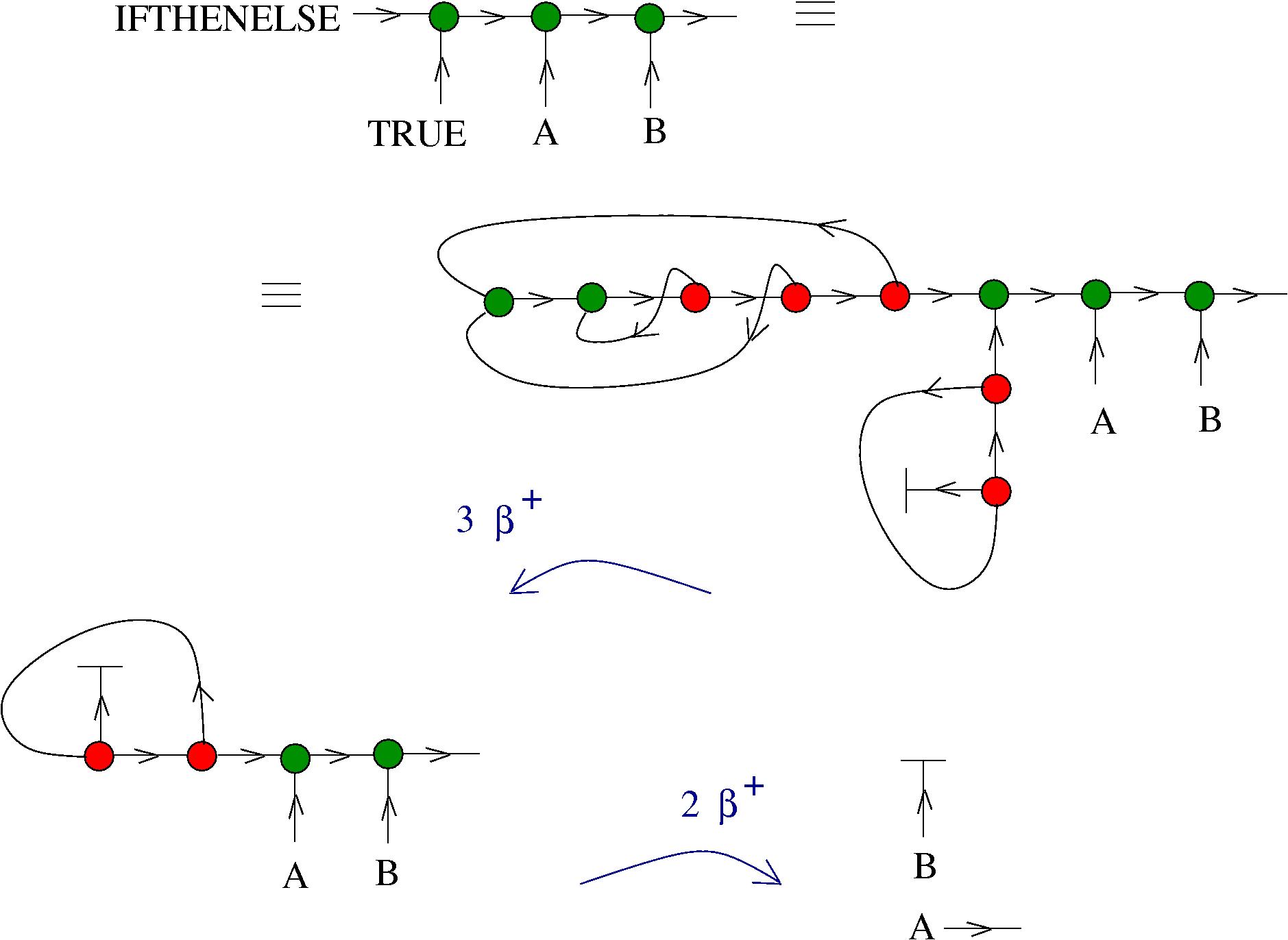}
}
\vspace{.5cm}

A rewrite which is introduced later in the article is DISENTANGLE  (as a pattern it appears later as SHUFFLE)

\vspace{.5cm}

\vspace{.5cm}
 
\centerline{\includegraphics[width=0.8\textwidth]{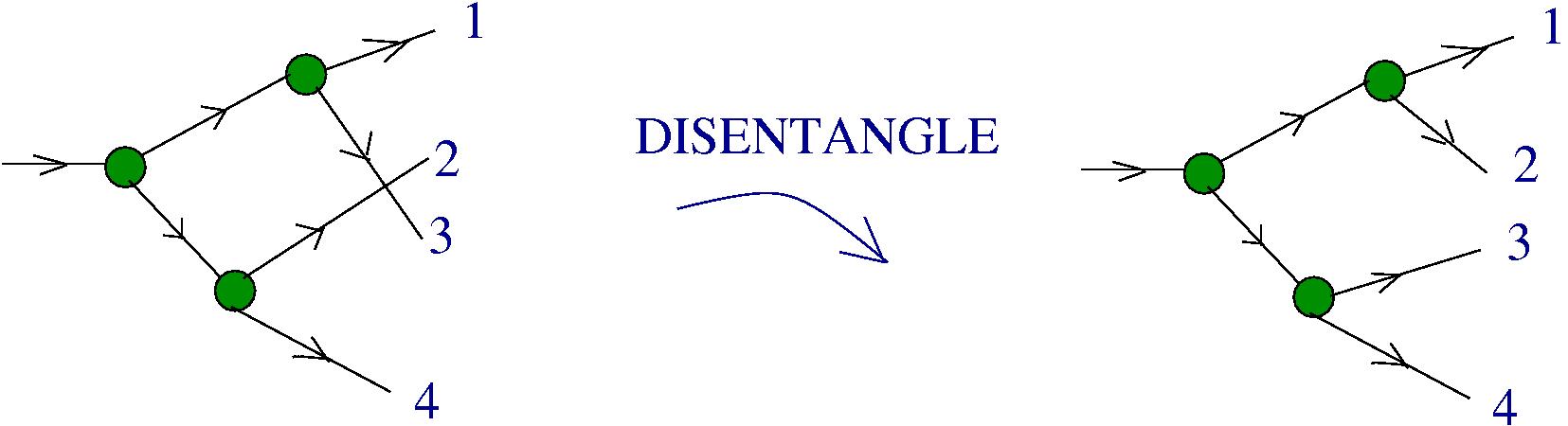}
}
\vspace{.5cm}

\vspace{.5cm}
With this supplementary rewrite  I can prove that the BCKW combinators system

\vspace{.5cm}
 
\centerline{\includegraphics[width=0.5\textwidth]{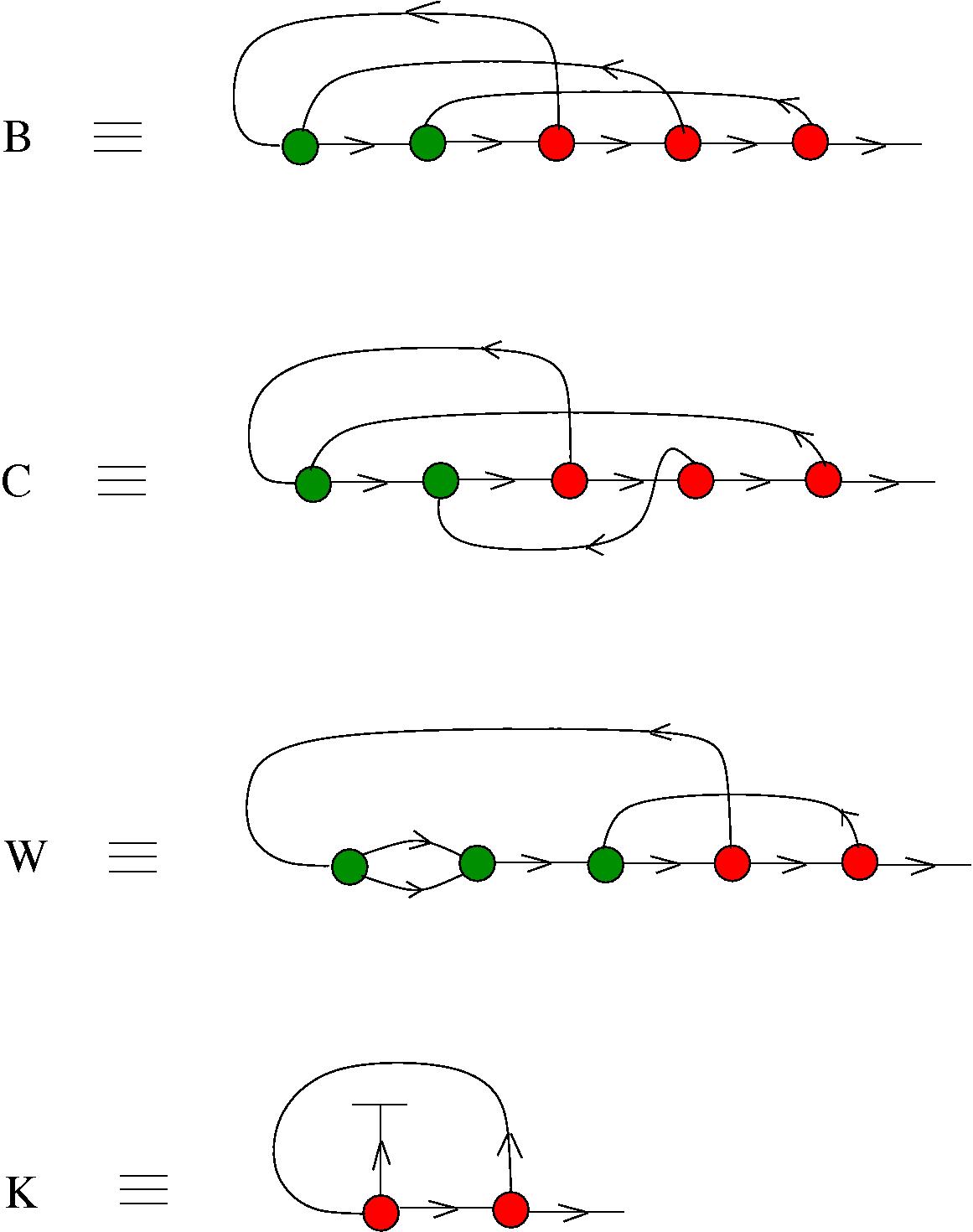}
}
\vspace{.5cm}

can be implemented, in the sense that the combinators can duplicate (example here for B)

\vspace{.5cm}
 
\centerline{\includegraphics[width=0.8\textwidth]{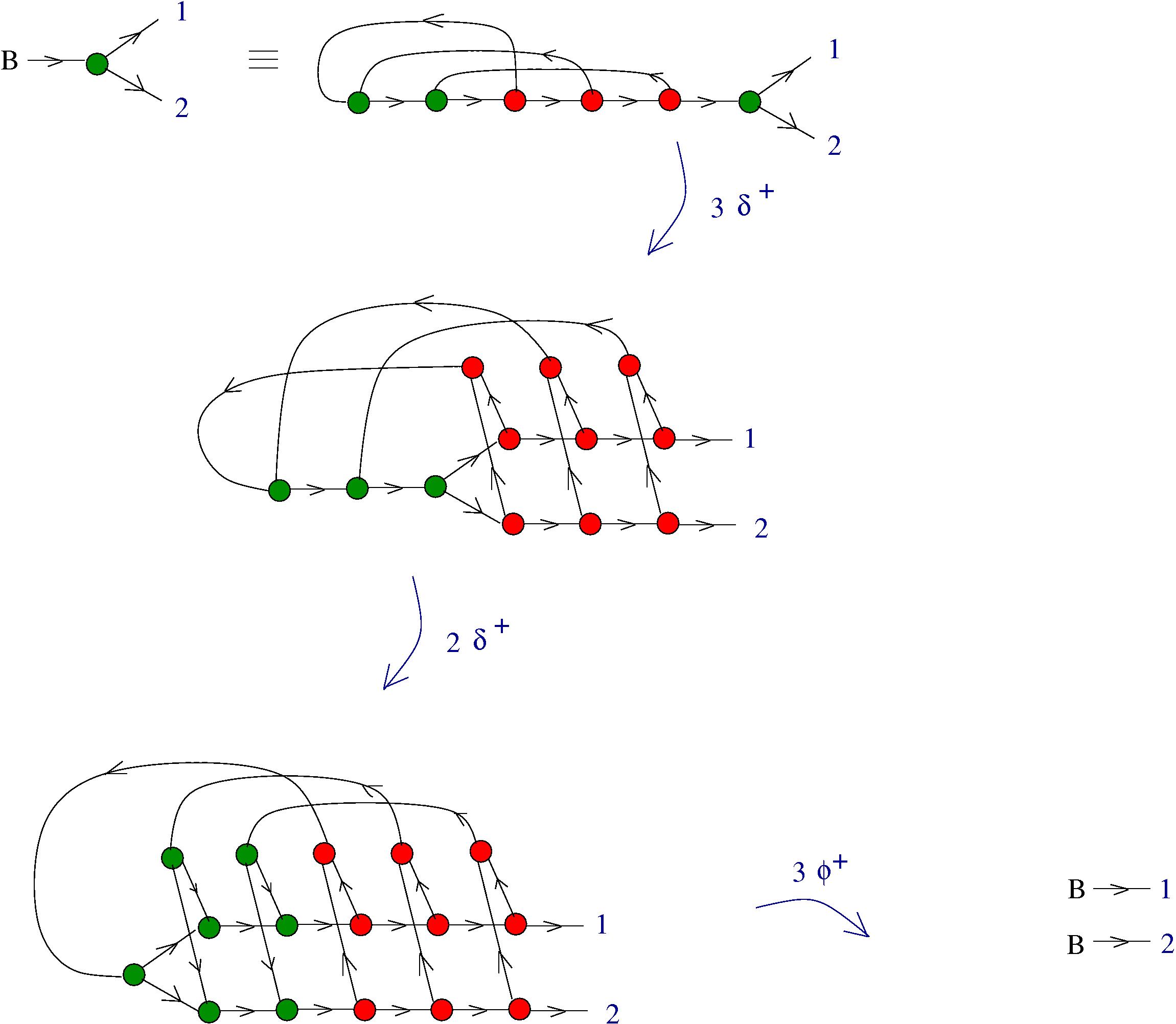}
}
\vspace{.5cm}

and that all needed reductions among B,C,K,W combinators can be done (example here for K)

\vspace{.5cm}
 
\centerline{\includegraphics[width=0.7\textwidth]{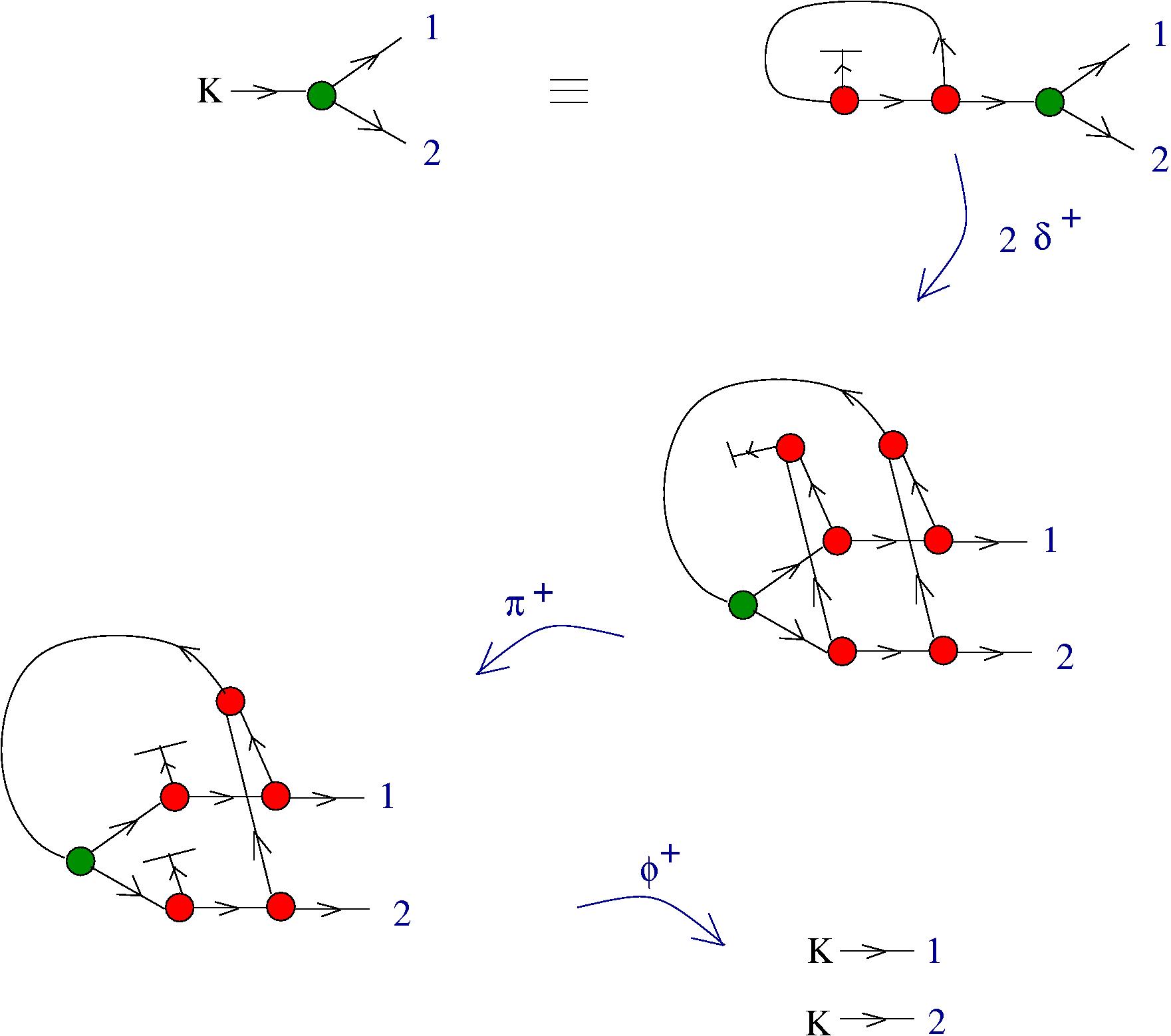}
}
\vspace{.5cm}

This graph rewrite system has conflicting rewrites, but it has no algorithm of application of rewrites. Especially the status of DISENTANGLE is unclear. This will be solved in chemlambda v2.

The first published (in the usual ways) article which contains the name "chemlambda" is joint with L.H. Kauffman, in the ALIFE-14 conference, \href{https://arxiv.org/abs/1403.8046}{arXiv:1403.8046}. Here Kauffman pushes again for a more topological version \href{https://mbuliga.github.io/emergent-10-years/pdf/ALIFETalk-copy.pdf}{Kauffman'  slides} \cite{kauffmanslides} which mixes tangle diagrams with chemlambda nodes, in order to be able to do computations (in the logical sense).The emergent algebras part of GLC, which is still present in chemlambda, used such tangle diagrams manipulations for differential calculus computations, instead. In these articles the emergent algebras aspects and interest are hidden.

Together with Kauffman, we experimented a lot with the power of chemlambda v1. Our essays on paper were not going too far, due to the lack of a program which could make much easier such exploration. In the background this was because of a lack of a clear rewrite application algorithm.

During this period I became aware of the existence of chemlambda quines. See as an example the tedious manipulation of reductions for the predecessor of a Church number taken from \href{https://chorasimilarity.wordpress.com/2014/08/27/ouroboros-predecessor-iii-the-walking-machine/}{this chorasimilarity post}
\vspace{.5cm}

\vspace{.5cm}
 
\centerline{\includegraphics[width=0.6\textwidth]{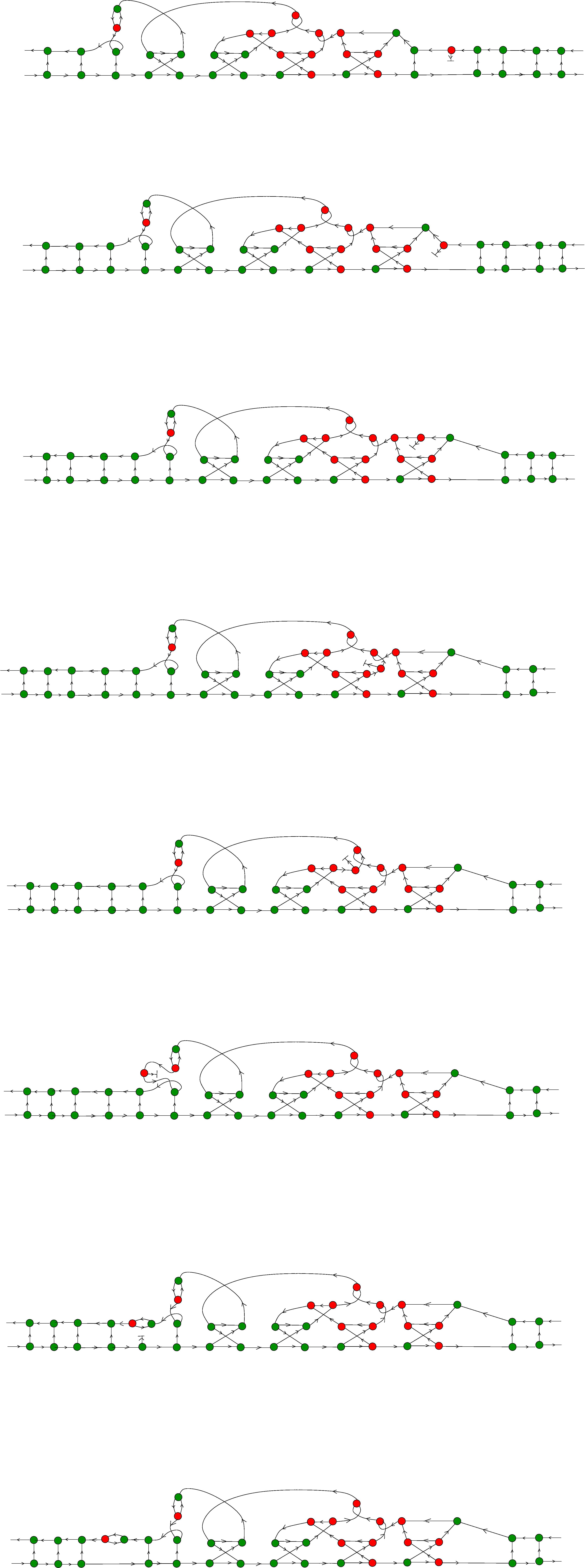}
}
\vspace{.5cm}
Meanwhile Kauffmann started to use Mathematica for doing chemlambda v1 reductions, as described in his slides \cite{kauffmanslides}.

\vspace{.5cm}
 
\centerline{\includegraphics[width=0.8\textwidth]{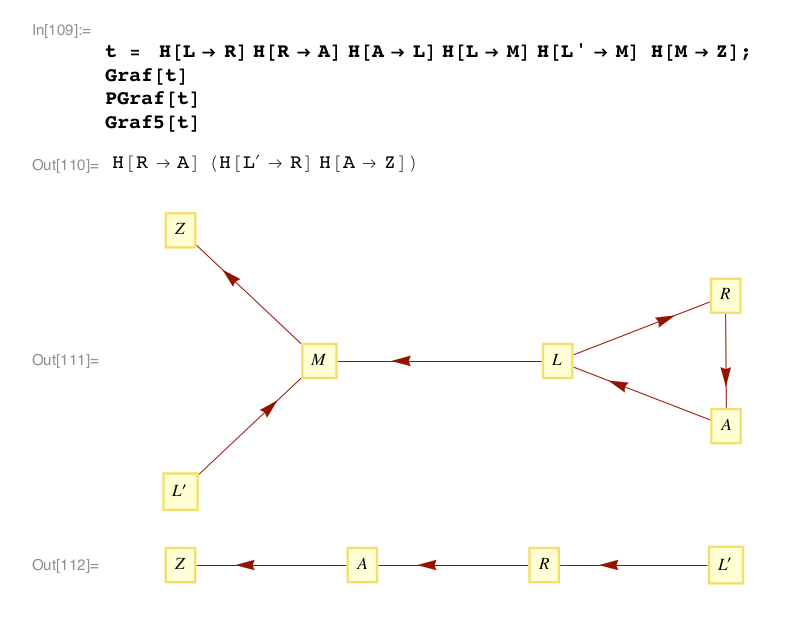}
}
He introduced the "Arrow" 2-valent node, which is useful for easy application of several rewrites in the same time.

\section{Chemlambda v2}
\label{ChemlambdaV2}
is an artificial chemistry, in the sense that it is a purely local graph rewrite system together with an algorithm of applications which can be done by local machines. This notion of artificial chemistry is proposed in \cite{tomita}. See \href{https://arxiv.org/abs/2005.06060}{arXiv:2005.06060} for more explanations.

Motivated by the first programming essays by Kauffman for chemlambda v1, I wrote the needed programs for  chemlambda v2, in awk and html+d3.js used for visualization. The \href{http://chorasimilarity.github.io/chemlambda-gui/index.html}{first chemlambda project site} \cite{first} contains demonstrations made with these programs from the github repository
\href{https://github.com/chorasimilarity/chemlambda-gui/blob/gh-pages/dynamic/README.md}{chemlambda-gui} \cite{firstrepo}.

The latest chemlambda project site, which contains several repositories from various contributors, is \href{https://chemlambda.github.io/index.html}{this} \cite{entry}. The project received help from several contributors, or validators, therefore now we can reduce chemlambda molecules in haskell, python or javascript, as you can see in the more recent demonstrations.

Chemlambda can now be \href{https://mbuliga.github.io/quinegraphs/ice.html#howto}{compared with Interaction Combinators} \cite{buligaalife} and there is a \href{https://mbuliga.github.io/quinegraphs/lambda2mol.html#lambdanote}{lambda calculus to chemlambda parser and reducer} \cite{lambda2mol}.

The chemlambda v2 rewrites solve the problem of the curious DISENTANGLE rewrite by introducing two fanout nodes (FO and FOE) instead of one and by a modified set of graph rewrites. These are the following. We use the mol notation for clarity and we describe the rewrites by giving the LHS and RHS patterns:

\vspace{.5cm}
 
\centerline{\includegraphics[width=0.5\textwidth]{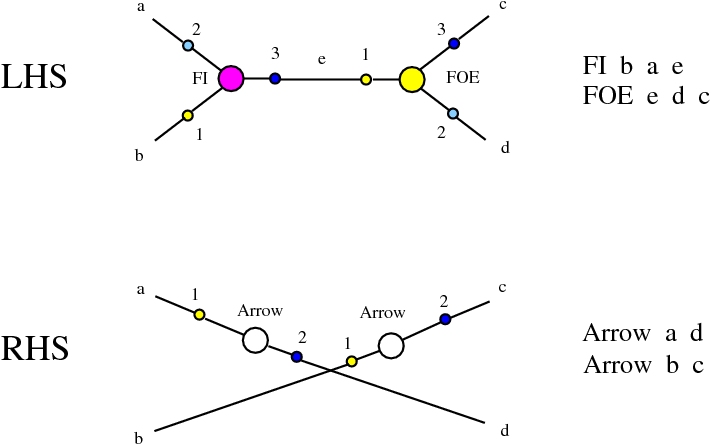}
}
\vspace{.5cm}
\centerline{\includegraphics[width=0.5\textwidth]{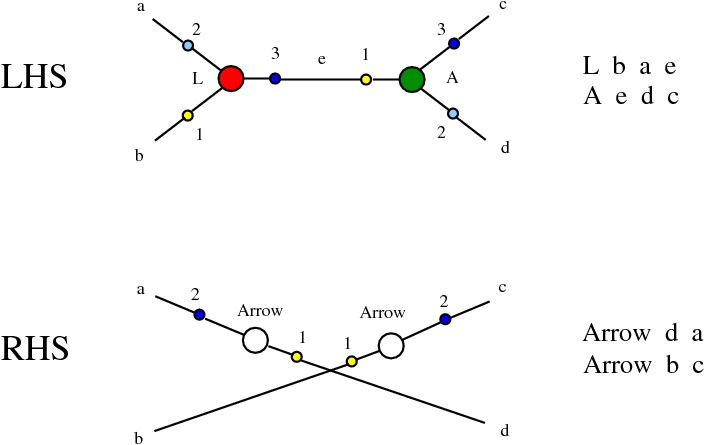}
}
\vspace{.5cm}

\vspace{.5cm}
 
\centerline{\includegraphics[width=0.5\textwidth]{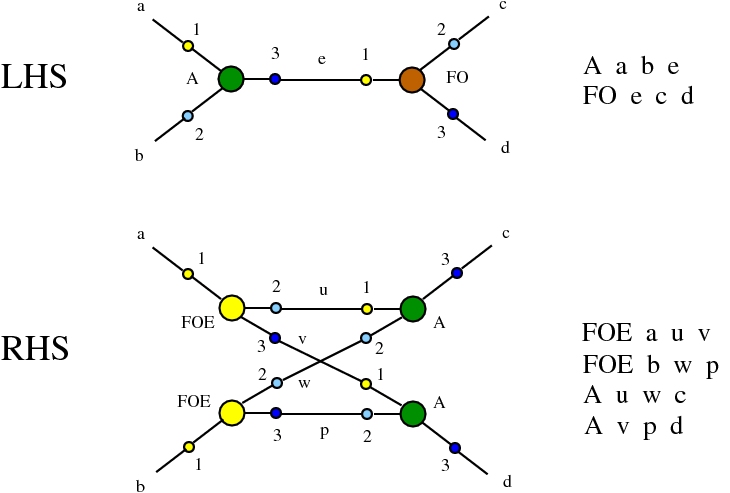}
}
\vspace{.5cm}

\centerline{\includegraphics[width=0.5\textwidth]{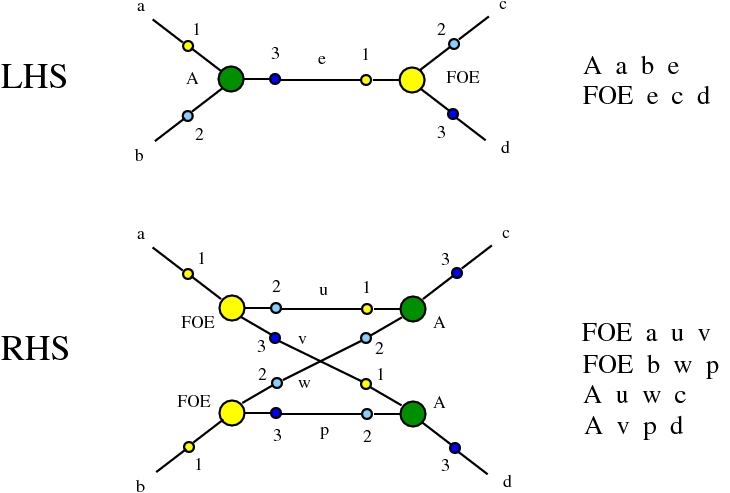}
}
\vspace{.5cm}

\vspace{.5cm}
 
\centerline{\includegraphics[width=0.5\textwidth]{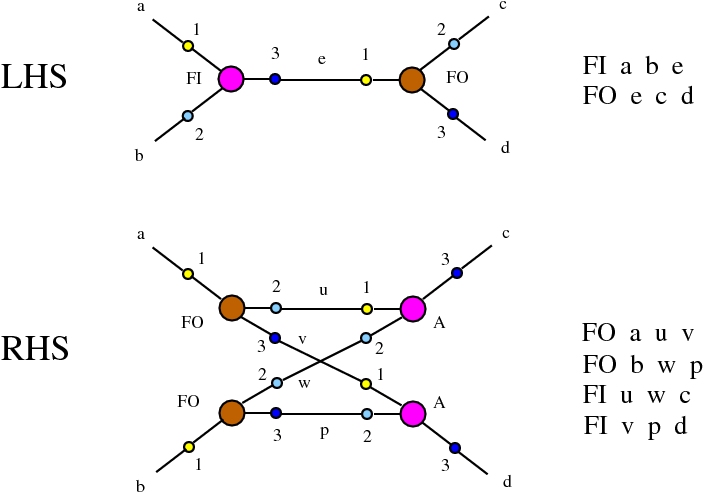}
}
\vspace{.5cm}

\centerline{\includegraphics[width=0.5\textwidth]{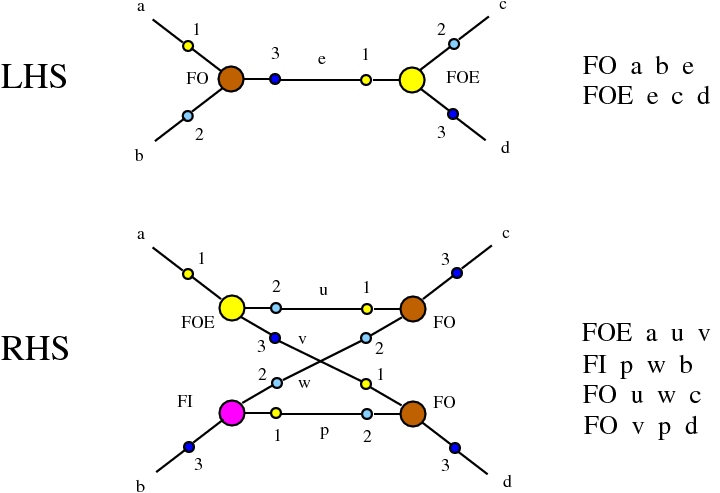}
}
\vspace{.5cm}

\vspace{.5cm}
 
\centerline{\includegraphics[width=0.5\textwidth]{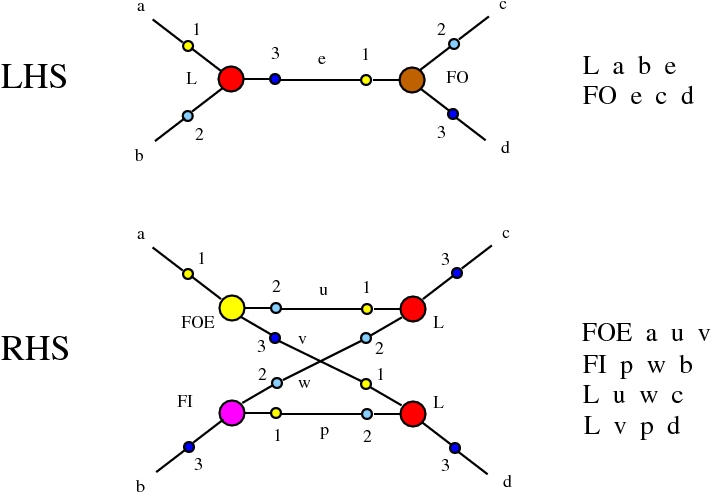}
}
\vspace{.5cm}

\centerline{\includegraphics[width=0.5\textwidth]{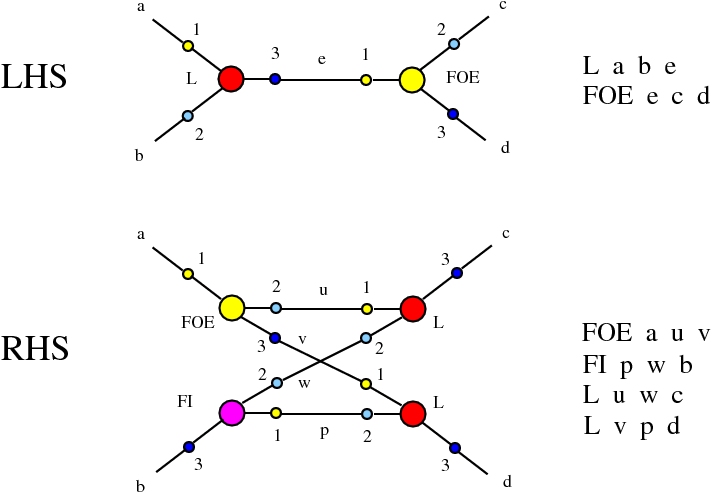}
}
\vspace{.5cm}

\vspace{.5cm}
 
\centerline{\includegraphics[width=0.5\textwidth]{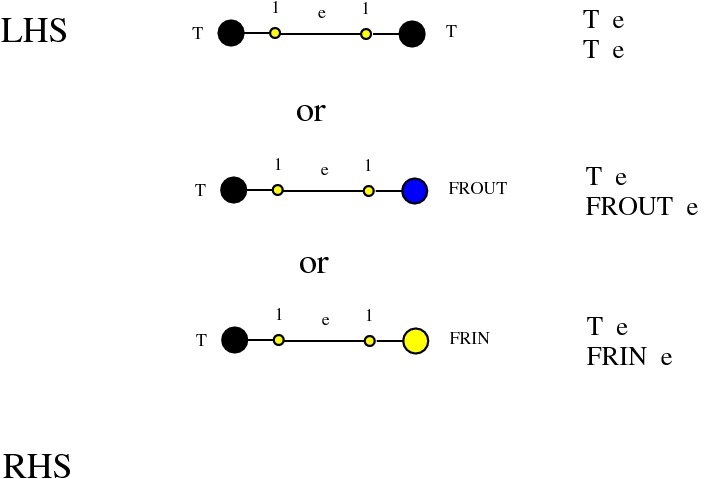}
}
\vspace{.5cm}

\centerline{\includegraphics[width=0.5\textwidth]{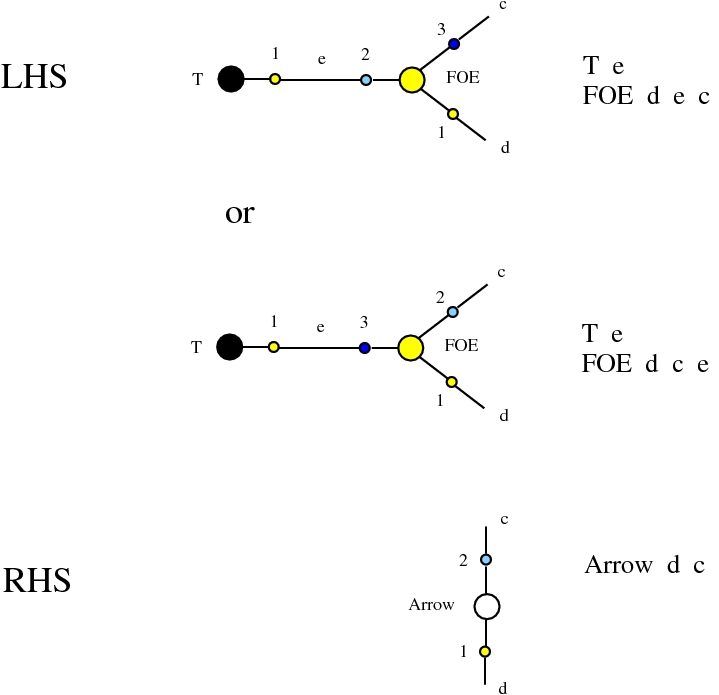}
}
\vspace{.5cm}

\vspace{.5cm}
 
\centerline{\includegraphics[width=0.5\textwidth]{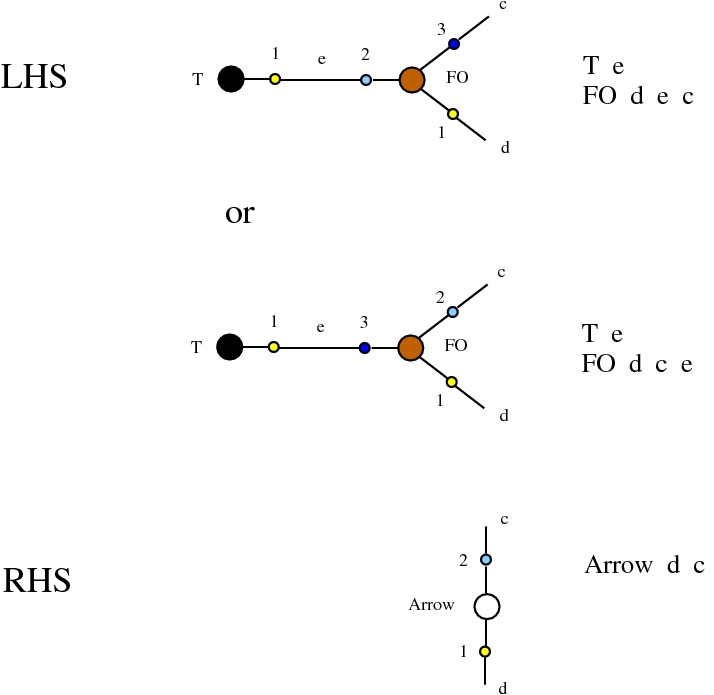}
}
\vspace{.5cm}

\centerline{\includegraphics[width=0.5\textwidth]{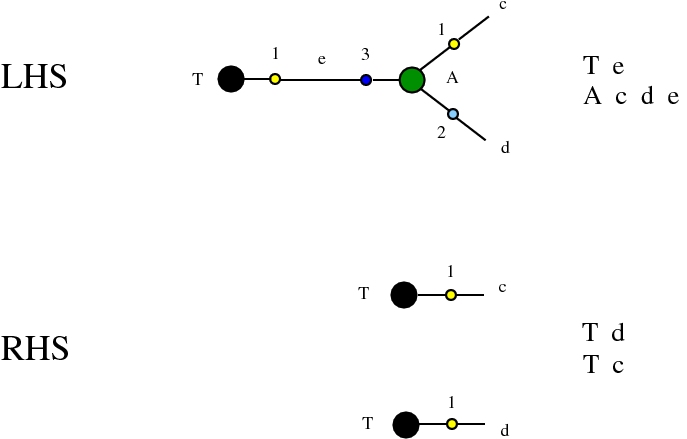}
}
\vspace{.5cm}

\vspace{.5cm}
 
\centerline{\includegraphics[width=0.5\textwidth]{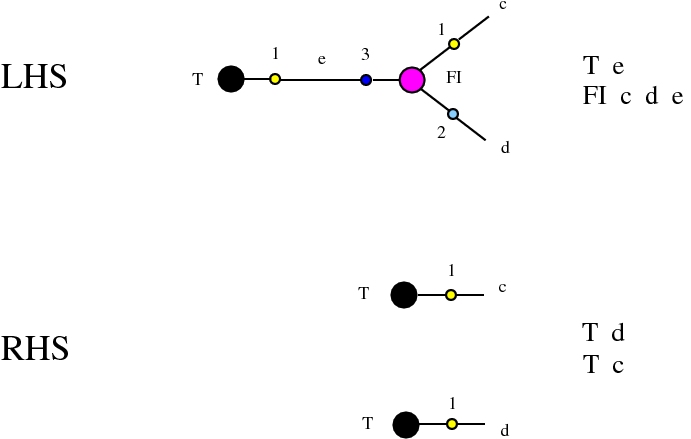}
}
\vspace{.5cm}

\centerline{\includegraphics[width=0.5\textwidth]{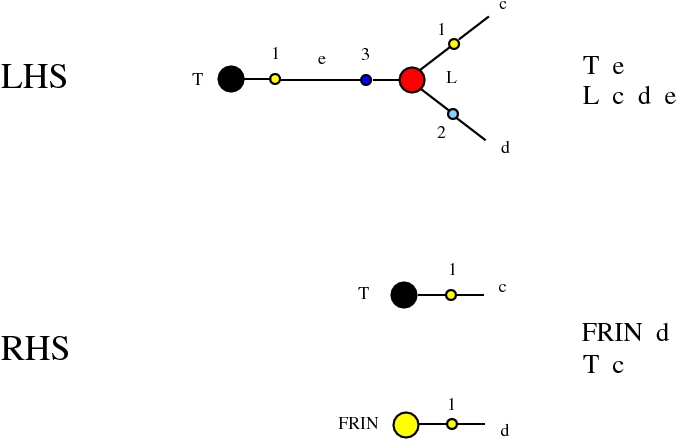}
}
\vspace{.5cm}

\vspace{.5cm}
 
\centerline{\includegraphics[width=0.5\textwidth]{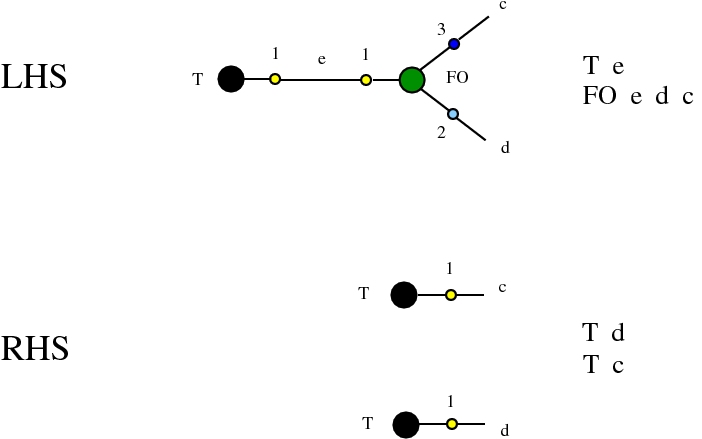}
}
\vspace{.5cm}

\centerline{\includegraphics[width=0.5\textwidth]{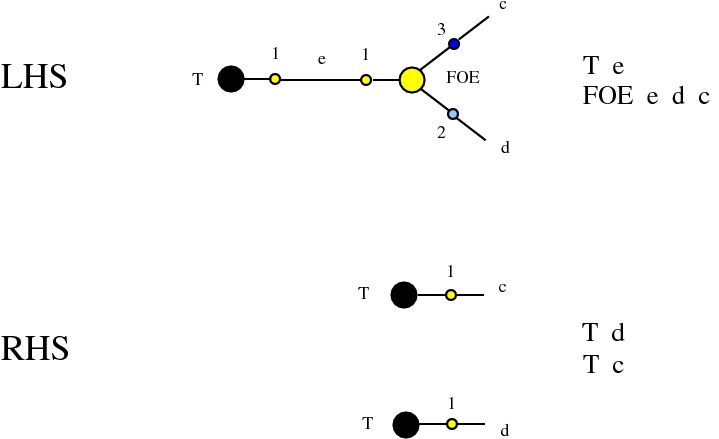}
}
\vspace{.5cm}

\vspace{.5cm}
 
\centerline{\includegraphics[width=0.5\textwidth]{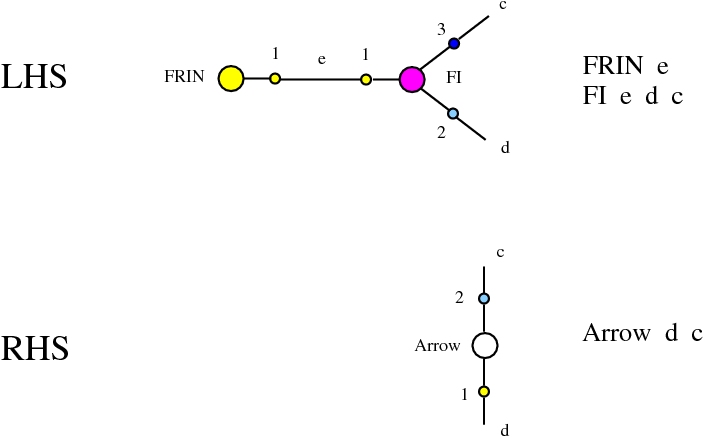}
}
\vspace{.5cm}

\centerline{\includegraphics[width=0.5\textwidth]{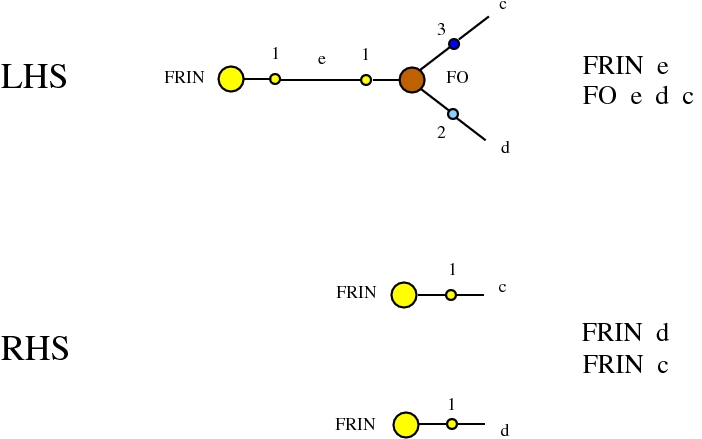}
}
\vspace{.5cm}

and the COMB rewrites which eliminate Arrow nodes.

The DISENTANGLE rewrite is no longer needed, because it's effect can be achieved by the SHUFFLE move, which is a sequence of two chemlambda v2 rewrites: FO-FOE and FI-FOE.

\vspace{.5cm}
 
\centerline{\includegraphics[width=0.5\textwidth]{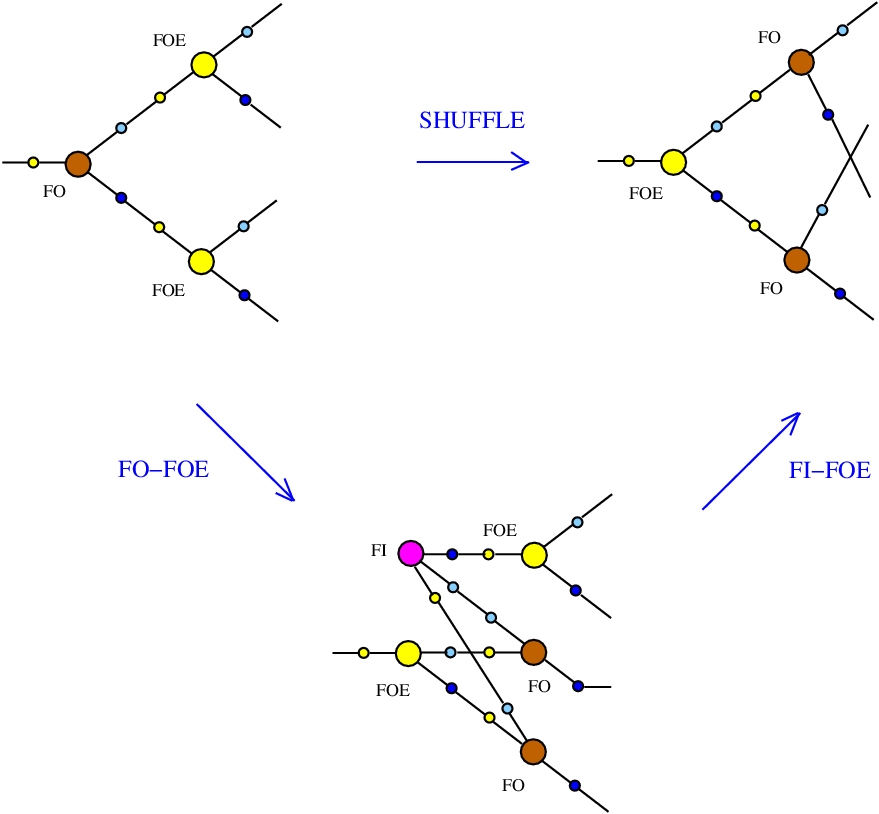}
}
\vspace{.5cm}

\section{Interaction Combinators and directed interaction combinators (dirIC)}
\label{IC}
\vspace{.5cm}
For completeness, here are the graph rewrites of Lafont' Interaction combinators \cite{lafont-comb}, as they appear in the chemlambda v2 js programs, chemistry IC in \href{https://github.com/mbuliga/quinegraphs/blob/master/js/chemistry.js}{chemistry.js}: 

\vspace{.5cm}
 
\centerline{\includegraphics[width=0.5\textwidth]{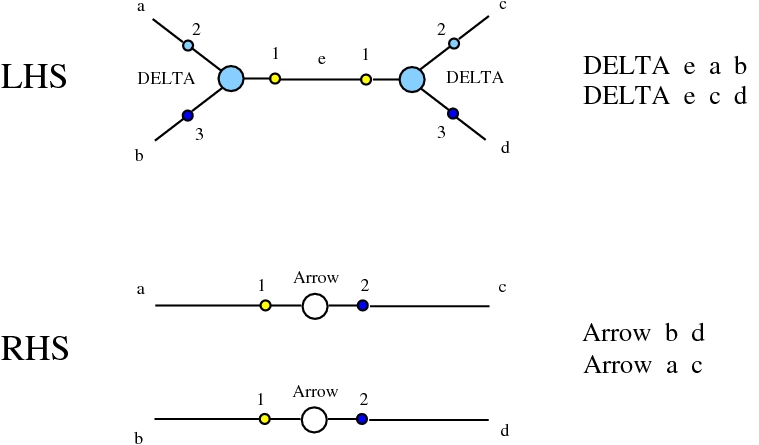}
}
\vspace{.5cm}

\centerline{\includegraphics[width=0.5\textwidth]{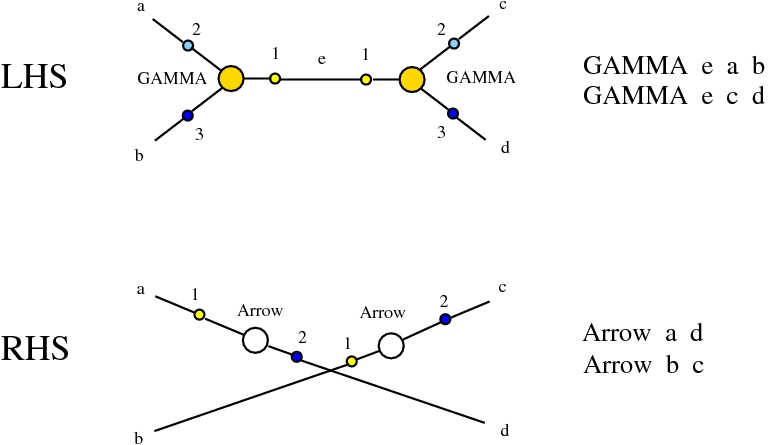}
}
\vspace{.5cm}

\vspace{.5cm}
 
\centerline{\includegraphics[width=0.5\textwidth]{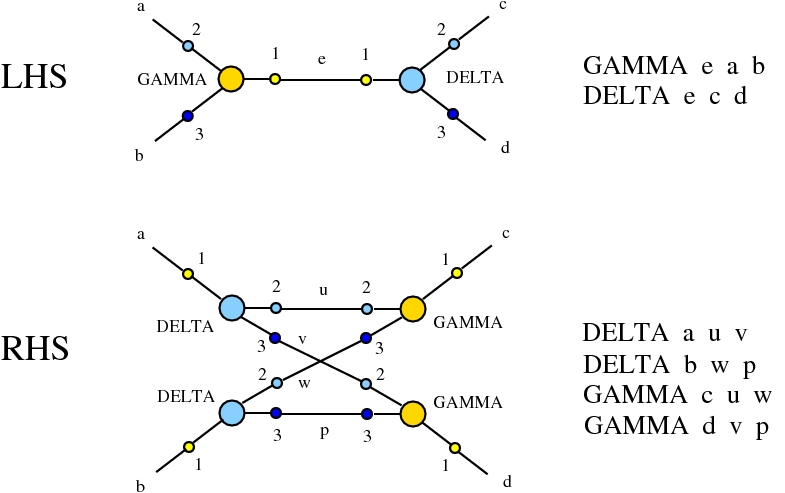}
}
\vspace{.5cm}

\vspace{.5cm}
 
\centerline{\includegraphics[width=0.5\textwidth]{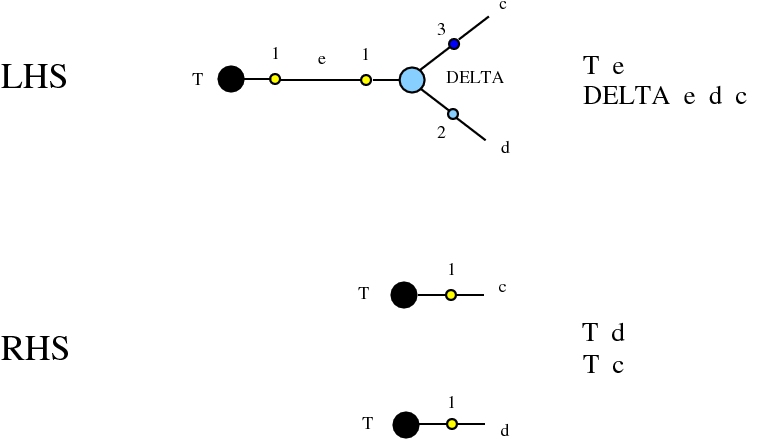}
}
\vspace{.5cm}

\centerline{\includegraphics[width=0.5\textwidth]{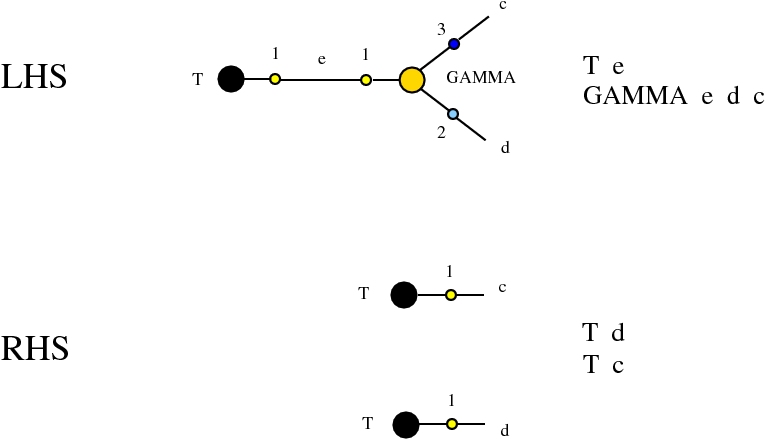}
}
\vspace{.5cm}

\vspace{.5cm}
 
\centerline{\includegraphics[width=0.5\textwidth]{remove4-T-T.jpg}
}
\vspace{.5cm}

FRIN and FROUT 1-valent nodes are introduced (as in chemlambda) for the free half-edges. We use also the Arrow 2-valent node and the COMB rewrites for eliminating them.

All LHS patterns, of chemlambda and IC, are made of two nodes connected through their active ports. The main difference between chemlambda v2 rewrites and IC rewrites is that several chemlambda nodes have two active ports, while IC nodes have only one active port.

It is possible to modify the chemlambda v2 chemistry such that there are no conflicts. We obtain a chemistry called "dirIC", or directed interaction combinators, explained in \cite{buligaalife} \href{https://mbuliga.github.io/quinegraphs/ic-vs-chem.html#icvschem}{Alife properties of directed interaction combinators vs. chemlambda}, over the same set of nodes as chemlambda v2. With the chemistry dirIC, there is a translation between IC nodes of Lafont to chemlambda nodes: 

\vspace{.5cm}
 
\centerline{\includegraphics[width=0.5\textwidth]{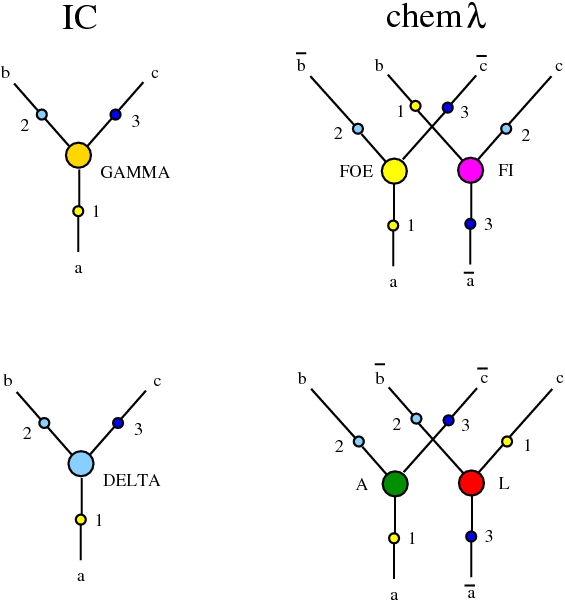}
}
\vspace{.5cm}

As argued \href{https://mbuliga.github.io/quinegraphs/ic-vs-chem.html#icvschem}{here} \cite{buligaalife}, the existence of conflicting rewrites in chemlambda v2, compared with dirIC, creates much more interesting alife behaviours.

\end{document}